\documentclass[sigconf]{acmart}

\copyrightyear{2022}
\acmYear{2022}
\setcopyright{acmcopyright}\acmConference[ASE '22]{37th IEEE/ACM International Conference on Automated Software Engineering}{October 10--14, 2022}{Rochester, MI, USA}
\acmBooktitle{37th IEEE/ACM International Conference on Automated Software Engineering (ASE '22), October 10--14, 2022, Rochester, MI, USA}
\acmPrice{15.00}
\acmDOI{10.1145/3551349.3561157}
\acmISBN{978-1-4503-9475-8/22/10}

\usepackage{bbding}

\usepackage{graphicx}
\usepackage{pgfplots}
\usepackage{pgfplotstable}
\usepackage{subcaption}
\usepackage{cellspace, makecell}
\usepackage[many]{tcolorbox}

\usepackage{dsfont}
\usepackage{mathtools,amssymb,latexsym,amsfonts,stmaryrd}
\usepackage{pbox}
\usepackage{xfrac}
\usepackage{booktabs}
\usepackage{xcolor,colortbl}
\usepackage{threeparttable}
\usepackage{amsthm}
\usepackage{scalerel}

\usepackage{hyperref}
\usepackage{multirow}
\usepackage{tikz}
\usepackage{mathrsfs}
\usepackage{mathpartir}
\usepackage[ruled,linesnumbered,vlined,noend]{algorithm2e}
\usepackage{caption}
\usepackage{url}
\usepackage{syntax}
\usepackage{framed}
\usepackage[noend]{algpseudocode}
\usepackage{flushend}
\usepackage{listings}
\usepackage{moresize}
\usepackage{wrapfig}
\def\checkmark{\tikz\fill[scale=0.4](0,.35) -- (.25,0) -- (1,.7) -- (.25,.15) -- cycle;} 
\usepackage{seqsplit}
\usepackage{alltt}
\usepackage{xspace}
\usepackage{enumitem}
\usepackage{pifont}

\DeclareMathAlphabet{\mathcal}{OMS}{cmsy}{m}{n}

\usepackage{amsmath, listings, amsthm, amssymb, proof, xspace}

\usepackage{array}
\newcommand{\PreserveBackslash}[1]{\let\temp=\\#1\let\\=\temp}
\newcolumntype{C}[1]{>{\PreserveBackslash\centering}p{#1}}
\newcolumntype{R}[1]{>{\PreserveBackslash\raggedleft}p{#1}}
\newcolumntype{L}[1]{>{\PreserveBackslash\raggedright}p{#1}}

\usepackage{thmtools}

\declaretheoremstyle[spaceabove=\topsep,notefont=\normalfont\itshape]{mystyle}

\newcommand{\revise}[2]{{\color{red}{\ifx&#1&\else- #1\fi}} {\color{ForestGreen}{\ifx&#2&\else+ #2\fi}}}%
\renewcommand{\revise}[2]{#2}%

\usetikzlibrary{arrows,patterns, decorations.pathreplacing}
\newtheorem{definition}{Definition}

\newcommand{\F}{Fig.}
\newcommand{\E}{Eq.}
\newcommand{\T}{Table}
\renewcommand{\S}{Sec.}

\newcommand{\ignore}[1]{}

\newcommand{\parh}[1]{\noindent\textbf{#1}}
\newcommand{\parhs}[1]{\noindent\underline{#1}}

\newcommand{\Df}{Definition}

\lstdefinestyle{base}{
  moredelim=**[is][\color{red}]{@}{@},
  escapeinside={<@}{@>}
}

\newcommand{\dl}{\textsc{DeepLIFT}}
\newcommand{\vcc}{visual concepts}

\newcommand{\mt}{$MR_{t}$}
\newcommand{\mr}{$MR_{r}$}

\newcommand{\synbracket}[1]{[\![#1]\!]}

\usepackage{amssymb}
\usepackage{pifont}

\newcommand\DejaVuttfamily{%
  \fontfamily{DejaVuSansMono-TLF}\selectfont }

\lstdefinestyle{base}{
  moredelim=**[is][\color{red}]{@}{@},
  escapeinside={<@}{@>}
}

\lstdefinelanguage
   [x64]{Assembler}     
   [x86masm]{Assembler} 
   {morekeywords={CDQE,CQO,CMPSQ,CMPXCHG16B,JRCXZ,LODSQ,MOVSXD, %
                  POPFQ,PUSHFQ,SCASQ,STOSQ,IRETQ,RDTSCP,SWAPGS, %
                  rax,rdx,rcx,rbx,rsi,rdi,rsp,rbp, %
                  r8,r8d,r8w,r8b,r9,r9d,r9w,r9b}} 

\usepackage[compact]{titlesec}

\usepackage{listings}
\usepackage{fancyvrb}
\usepackage{color}
\definecolor{lightgray}{rgb}{.9,.9,.9}
\definecolor{darkgray}{rgb}{.4,.4,.4}
\definecolor{purple}{rgb}{0.65, 0.12, 0.82}
\definecolor{commentgreen}{RGB}{63,127,95}

\colorlet{myPurple}{blue!40!red}
\definecolor{myOrange}{RGB}{255,192,0}

\newcommand{\enc}[1]{$\phi^{*}_{\theta}$}
\newcommand{\dec}[1]{$\psi^{*}_{\theta}$}

\lstdefinelanguage{Solidity}{
  keywords={len,delete,int,void,payable, public, event, contract, typeof, new, true, false, catch, function, return, null, catch, switch, var, if, in, while, do, else, case, break,struct,const,socklen_t,sa_familty_t,char,sockaddr},
  keywordstyle=\color{violet}\bfseries,
  ndkeywords={class, export, boolean, throw, implements, import, this},
  ndkeywordstyle=\color{darkgray}\bfseries,
  identifierstyle=\color{black},
  sensitive=false,
  comment=[l]{//},
  escapeinside={(*@}{@*)},          
  morecomment=[s]{/*}{*/},
  commentstyle=\color{commentgreen}\ttfamily,
  stringstyle=\color{red}\ttfamily,
  morestring=[b]',
  morestring=[b]"
}

\newcommand{\rnum}[1]{\uppercase\expandafter{\romannumeral #1\relax}}

\usepackage{xcolor}

\algnewcommand{\LeftComment}[1]{\Statex \(\triangleright\) #1}

\definecolor{pptbrown}{RGB}{132,60,12}
\definecolor{pptgreen}{RGB}{56,87,35}

\let\OLDthebibliography\thebibliography
\renewcommand\thebibliography[1]{
  \OLDthebibliography{#1}
  \setlength{\parskip}{0pt}
  \setlength{\itemsep}{0pt plus 0.1ex}
}

\definecolor{pptgreen}{RGB}{84,130,53}
\definecolor{pptred}{RGB}{176,35,24}

\definecolor{pptgreen1}{RGB}{78,173,91}
\definecolor{pptred1}{RGB}{192,0,0}

\definecolor{pptyellow1}{RGB}{203,195,167}
\definecolor{pptgreen2}{RGB}{184,192,176}

\definecolor{pptred3}{RGB}{192,0,0}
\definecolor{pptyellow3}{RGB}{255,192,0}
\definecolor{pptgreen3}{RGB}{4,216,178}

\definecolor{pptblue}{RGB}{0,176,240}
\definecolor{pptgrey}{RGB}{175,171,171}

\newcommand{\CBrush}{\textcolor{green}{\checkmark}}
\newcommand{\XBrush}{\textcolor{red}{\ding{55}}}

\newlength{\dpcircle}
\newlength{\rcircle}
\newlength{\dcircle}
\newcommand{\docircle}[4]{%
  \setlength{\dpcircle}{\dp\strutbox}%
  \setlength{\dcircle}{\dpcircle}%
  \addtolength{\dcircle}{\ht\strutbox}%
  \setlength{\rcircle}{0.5\dcircle}%
  \setlength{\unitlength}{1sp}%
  \begin{picture}(\number\dcircle,0)
    \color{#1}
    \put(\number\rcircle,\number\dpcircle){\circle*{\number\dcircle}}
    \color{#2}
    \put(\number\rcircle,\number\dpcircle){\circle{\number\dcircle}}
    \put(\number\rcircle,0){\makebox[0pt]{\textcolor{#3}{#4}}}
  \end{picture}%
}

\newcommand{\CircleOne}{\ding{172}}
\newcommand{\CircleTwo}{\ding{173}}
\newcommand{\CircleThree}{\ding{174}}
\newcommand{\CircleFour}{\ding{175}}

\newcommand{\Circ}[1]{{\footnotesize\docircle{white}{black}{black}{#1}}}

\newcommand\rowincludegraphics[2][]{\raisebox{-0.45\height}{\includegraphics[#1]{#2}}}

\begin{document}

\title{Unveiling Hidden DNN Defects with Decision-Based Metamorphic Testing}
\titlenote{The extended version of the ASE 2022 paper~\cite{yuan2022unveiling}.}

\author{Yuanyuan Yuan}
\affiliation{%
  \institution{The Hong Kong University of Science and Technology}
  \country{Hong Kong, China}
}
\email{yyuanaq@cse.ust.hk}

\author{Qi Pang}
\affiliation{%
  \institution{The Hong Kong University of Science and Technology}
  \country{Hong Kong, China}
}
\email{qpangaa@cse.ust.hk}

\author{Shuai Wang}
 \affiliation{%
  \institution{The Hong Kong University of Science and Technology}
  \country{Hong Kong, China}
 }
\authornote{Corresponding Author}
\email{shuaiw@cse.ust.hk}

\begin{abstract}

    Contemporary DNN testing works are frequently conducted using metamorphic
    testing (MT). In general, de facto MT frameworks mutate DNN input images
    using semantics-preserving mutations and determine if DNNs can yield
    consistent predictions. Nevertheless, we find that DNNs may \textit{rely on
    erroneous decisions (certain components on the DNN inputs) to make
    predictions}, which may still retain the outputs by chance. Such DNN defects
    would be neglected by existing MT frameworks. Erroneous decisions, however,
    would likely result in successive mis-predictions over diverse images that
    may exist in real-life scenarios.

    This research aims to unveil the pervasiveness of hidden DNN defects caused
    by incorrect DNN decisions (but retaining consistent DNN predictions). To do
    so, we tailor and optimize modern eXplainable AI (XAI) techniques to
    identify visual concepts that represent regions in an input image upon which
    the DNN makes predictions. Then, we extend existing MT-based DNN testing
    frameworks to check the \textit{consistency of DNN decisions} made over a
    test input and its mutated inputs. Our evaluation shows that existing MT
    frameworks are oblivious to a considerable number of DNN defects caused by
    erroneous decisions. We conduct human evaluations to justify the validity of
    our findings and to elucidate their characteristics. Through the lens of DNN
    decision-based metamorphic relations, we re-examine the effectiveness of
    metamorphic transformations proposed by existing MT frameworks. We summarize
    lessons from this study, which can provide insights and guidelines for
    future DNN testing.

\end{abstract}



\maketitle

\section{Introduction}
\label{sec:introduciton}

Metamorphic testing (MT)~\cite{chen1998metamorphic} has achieved a major success
to comprehensively test deep neural networks (DNNs) without manually annotating
test inputs~\cite{ma2018deepmutation}. Given the inherent difficulty of defining
explicit testing oracles for DNN models~\cite{zhang2020machine}, DNN is often
tested using well-designed metamorphic relations (MRs): DNN inputs are mutated
into new test cases in a semantics-preserving manner\footnote{In this paper,
semantics-preserving denotes that the contents in inputs and mutated inputs 
are visually consistent, e.g., a cat is still a cat.}, and
DNN predictions over an input and its mutated inputs are compared for
consistency. DNN defects are characterized as violations of DNN prediction
consistency. However, despite the major success of checking prediction
consistency, we pose the following key question to motivate this research:

\begin{quote}
  ``\textit{Is it always the case that a consistent prediction indicates no DNN
  defect?}''
\end{quote}

\begin{figure}[!ht]
  \centering
  \includegraphics[width=1.0\linewidth]{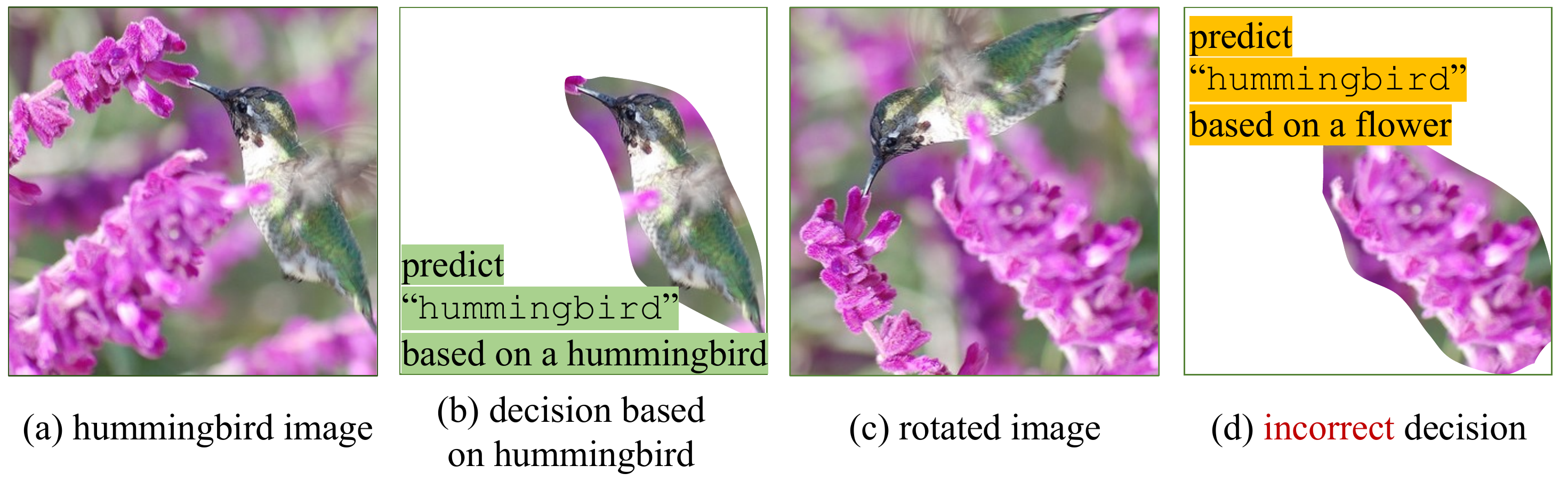}
  \vspace{-20pt}
  \caption{DNN is making inconsistent, erroneous decisions while happening to
  retain the same prediction. We simplify the decision regions for readability.}
  \vspace{-10pt}
  \label{fig:motivation}
\end{figure}

In this research, we refer to the DNN's focus on critical input components 
as its \textit{decisions}. Accordingly, DNN relies on such decisions to make
\textit{predictions} (i.e., its outputs), e.g., classifying an input image. 
Then, consider \F~\ref{fig:motivation}, in which we illustrate how a contemporary MT
framework misses a DNN defect. The tested DNN predicts ``hummingbird'' for
\F~\hyperref[fig:motivation]{1(a)}, and its utilized decisions in
\F~\hyperref[fig:motivation]{1(a)} are marked in
\F~\hyperref[fig:motivation]{1(b)}, depicting the correct scope of a
hummingbird's head and body. When \F~\hyperref[fig:motivation]{1(a)} is rotated
as in \F~\hyperref[fig:motivation]{1(c)}, the DNN still predicts
``hummingbird.'' Thus, existing MRs based on DNN output consistency would regard
the DNN as ``correct'' for this case. Nevertheless, as in
\F~\hyperref[fig:motivation]{1(d)}, the underlying DNN decision is
\textit{specious}, as it is based on a flower whose contour is similar to the
contour of the flying hummingbird in \F~\hyperref[fig:motivation]{1(b)}.

Our preliminary study shows that existing MT-based DNN testing frameworks, when
only checking the consistency of DNN predictions, may overlook DNN defects due
to incorrect DNN decisions, i.e., relying on specious components in the DNN
inputs for predictions. As revealed in this research, such incorrect decisions
do not always result in inconsistent DNN outputs, especially when the DNN is
trained on a dataset with limited labels (e.g., two-label classification),
because DNN prediction is forced to choose among pre-defined labels.
Consider a DNN $\phi$ that is trained on evenly distributed data and is performing
a two-class (cat vs. dog) classification task. Assume that when random noise is
applied to an image $i$, $\phi$ ignores the cat in $i$ and randomly guesses a
label. That is, it still has a 50\% chance of predicting the correct label.
Despite the fact that the tested DNN is flawed, it is nonetheless considered as
``robust to noise'' in many of these cases.

Moreover, we clarify that specious DNN decisions are hard to detect using only
the DNN predictions and confidence scores. A well-trained DNN predicts a
pre-defined label with a confidence score that is typically much higher than the
other labels, even when given random noise as inputs. Thus, even if DNN is
generating erroneous decisions, its outputs and accompanying confidence scores
often lack an evident ``pattern.'' This difficulty is also highlighted in prior
literatures~\cite{nguyen2015deep,hein2019relu}.

Overall, this work deems inconsistent DNN decisions exposed by MT are specious
and undesirable, as it is likely that a DNN, even if it happens to correctly
label an image (\F~\hyperref[fig:motivation]{1(c)}), will eventually mis-predict
a hummingbird for pervasive images existing in real-world scenarios. As
a result, we argue that failing to account for DNN decision defects may
jeopardize the reliability of present MT frameworks. We advocate that
\textit{proper MRs should take DNN decisions into consideration, rather than
merely checking DNN predictions}.

This work advocates to extend DNN prediction-based consistency checking, which
is extensively used in current MT, with decision-based consistency
checking. The enhancement is orthogonal to particular metamorphic
transformations (e.g., image pixel or affine transformations) implemented in
existing MT-based DNN testing frameworks, and can be smoothly incorporated by
them. Given an test image $i$, we extract the decision, denoting
regions in $i$, to depict how DNN makes prediction over $i$. Each region is
referred to as a visual concept (e.g., a nose or a wheel), and DNN predictions
can be formulated as a voting scheme among visual
concepts~\cite{linardatos2020explainable,gunning2019xai}.
To obtain visual concepts, we first use eXplainable AI (XAI) techniques to identify
pixels in $i$ that positively contribute to the DNN prediction. Then, we
tailor and optimize a set of image processing techniques to construct visual
concepts from XAI-identified pixels. We carefully reduce inherent inaccuracy of
XAI techniques, and largely enhance the readability of identified visual concepts.

By extending existing MT frameworks to support decision-based consistency
checking, we uncover many overlooked defects triggered by inputs that result in
inconsistent decisions but identical predictions. Our findings are justified by
large-scale and comprehensive (in total 10,000) human evaluations, where the
participants are 15 Ph.D. students having research experiences related to
DNNs and 10 other Ph.D. and masters students of various backgrounds.
Our study encompasses ten DNNs over three datasets of different scales and types
(e.g., RGB and black-white images) which are all popular in daily usage and have
been extensively tested by previous DNN testing research.
We summarize key lessons of this research, illustrating
that existing MT, when only checking DNN prediction consistency, may over-estimate
the reliability of DNNs. We also assess the strength of metamorphic
transformations (e.g., pixel mutation vs. adversarial perturbation) proposed by
existing work through the lens of our novel DNN decision view. Our findings and
summarized lessons can provide insights for follow-up enhancement of DNN
testing. In sum, we make the following contributions.

\begin{itemize}
  \item 
  We advocate that existing MT-based DNN testing should consider how
  DNN makes decisions rather than merely checking predictions. Accordingly, we 
  extend existing MRs by checking decision consistency to reveal DNN
  defects overlooked by existing works.
  
  \item Technically, we recast a DNN prediction as the outcome of a voting
  process among visual concepts in an input. We tailor and optimize image
  processing schemes to summarize visual concepts from image pixels positively
  contributing DNN predictions.
  
  \item Our study and human evaluation illustrate that many defects have been
  overlooked when only checking DNN prediction consistency. Our findings provide
  guidelines for users to calibrate MT-based DNN testing results, and also
  highlight further improvements that can be made by DNN testing.
  
\end{itemize}

\noindent \textbf{Artifact Availability.} To support results verification and
follow-up research comparison, we released code, data, and supplementary materials 
at~\url{https://github.com/Yuanyuan-Yuan/Decision-Oracle}~\cite{snapshot}.
\section{Preliminary and Motivation}
\label{sec:preliminary}

\subsection{Metamorphic Testing}
\label{subsec:mt}

DNNs are typically used to answer unknown questions where they are anticipated
to behave similarly to humans~\cite{zhang2020machine}. Given the diversity of
possible inputs encountered in real-life scenarios, obtaining ground truth
predictions in advance to assess DNN correctness is difficult, if not
impossible. Furthermore, even human experts may disagree on expected outputs of
certain edge cases. 

MT is extensively employed to test DNNs without the need for ground-truth or
explicitly defined testing oracles~\cite{chen1998metamorphic}. Overall,
each MR in MT composes a metamorphic transformation \mt\ and a relation \mr:
each \mt\ specifies a mutation scheme over a source input to generate a
follow-up test input, and the associated \mr\ defines the relationship of
expected outputs over the source and the mutated 
input~\cite{segura2016survey}. For instance, to test $sin(x)$, we can construct
an MR such that its \mt\ mutates an input $x$ into $\pi - x$, and the \mr\
checks the the equality relation $sin(x) = sin(\pi - x)$. In real-world usage,
\mr\ usually denotes invariant program properties. \mr\ should always hold when
arbitrarily mutating $x$ using \mt, and a bug in $sin(x)$ is detected whenever
\mr\ is violated. 

MT achieves major success in testing DNN models and
infrastructures~\cite{ma2020metamorphic,ma2021mt,xiao2022metamorphic,yuan2021enhancing,li2022cctest,
wang2020metamorphic,tian2018deeptest,xie2018coverage,yuan2021perception,
demir2019deepsmartfuzzer,zhang2018deeproad,
dwarakanath2018identifying,nakajima2019generating,dwarakanath2019metamorphic}.
Given DNN inputs are often images, MRs in this field are often constructed to
perform lightweight, semantics-preserving (visually consistent) image mutations
\mt\ from different angles (see \S~\ref{sec:implementation} for a literature
review of \mt\ designed in previous works). \mr\ is defined in a simple and
unified manner such that DNN predictions should be \textit{consistent} over an
input image and its follow-up image generated by using \mt. Thus, violation of
\mr, denoting inconsistent DNN predictions, are DNN defects.

\begin{table}[ht!]
  \vspace{-5pt}
  \caption{Four \mr\ based on DNN decisions ($D_1,D_2$) and predictions
   ($L_1,L_2$) over an input and its mutated input.}
  \vspace{-10pt}
  \label{tab:decision}
  \centering
\resizebox{0.8\linewidth}{!}{
  \begin{tabular}{l|c|c|c|c}
    \hline
      & \CircleOne $D_1 = D_2$ & \CircleTwo $D_1 \neq D_2$ & \CircleThree $D_1 \neq D_2$ & \CircleFour $D_1 = D_2$  \\
      & $L_1 = L_2$ &  $L_1 \neq L_2$  & $L_1 = L_2$ & $L_1 \neq L_2$ \\
    \hline
    No defect? & \CBrush &  \XBrush     & \XBrush & NA \\
    \hline
  \end{tabular}
  }
\end{table}

\subsection{Forming \mr\ with DNN Decisions}
\label{subsec:oracle}

Without knowledge of a DNN's decision procedure, we argue that relying merely on
its output (as how existing \mr\ is formed) may result in the omission of some
defects. Given a pair of inputs $i_1$ and $i_2$ ($i_2$ is mutated from $i_1$
using a \mt), suppose the DNN yields prediction $L_1$ based on decision $D_1$,
yielding $L_2$ based on decision $D_2$.\footnote{$D$ is formed by identifying
DNN's decision over the input $i$; see \S~\ref{sec:approach} for details.}
Then, we have four combinations of decisions/predictions, as in
\T~\ref{tab:decision}. \CircleOne\ denotes a correct prediction (from the
perspective of MT), whereas \CircleTwo\ represents that the DNN provides
inconsistent predictions $L1 \neq L2$. As introduced in \S~\ref{subsec:mt},
existing MT frameworks rely on \CircleTwo\ to form \mr, and we clarify that
\CircleFour\ is not feasible: $D_1 = D_2 \leadsto L_1 \neq L_2$ violates the
nature of a DNN. 

We explore a new focus to form \mr, as in \CircleThree, where DNNs make
inconsistent decisions ($D_1 \neq D_2$), but still happen to retain the same
prediction ($L_1 = L_2$). We deem them as \textit{hidden DNN defects} that
are incorrectly overlooked by existing works.
Suppose a DNN $\phi$ answers if hummingbirds appear in an image. $\phi$ is
trained on a biased dataset where all hummingbirds hover in the air, and
therefore, $\phi$ wrongly relies on ``vertical objects'' to recognize
hummingbird. For the image in \F~\hyperref[fig:motivation]{1(a)}, it is properly
predicted by $\phi$ as ``yes'' due to the hovering hummingbird. After rotating
this image for 90 degrees as in \F~\hyperref[fig:motivation]{1(c)}, we find that
$\phi$ still responds ``yes,'' but makes decision based the vertically presented
flower in \F~\hyperref[fig:motivation]{1(d)}, which shares a similar contour to
most hummingbirds (e.g., by comparing with the contour in
\F~\hyperref[fig:motivation]{1(b)}). In fact, we manually retain only the visual
concept in \F~\hyperref[fig:motivation]{1(d)}, and erase the remaining
components of the image. We confirm that $\phi$ predicts the image as a
``hummingbird.''
Moreover, while $\phi$ is obviously susceptible to rotation, \mr\ based on
\CircleTwo\ cannot uncover the defect. Nevertheless, \mr\ based on \CircleThree\
can unveil this hidden flaw.

\parh{Paper Structure.}~In the rest of this paper, we formulate $D$ in
 \S~\ref{subsec:decision}, and present technical solutions to constitute $D$ in
 \S~\ref{sec:approach}. We review literatures of MT-based DNN testing and
 their proposed \mt\ in \S~\ref{sec:implementation}. \S~\ref{sec:evaluation}
 unveils the pervasiveness of hidden defects falling in \CircleThree\ with
 empirical results. 

\parh{Incompaliance of Ground Truth.}~Following the notation above, let the
ground truth prediction be $L_{G}$. It is widely seen that MT may result in
false negatives due to $L_{G} \neq (L_1 = L_2)$. That is, a DNN makes consistent
albeit incorrect predictions over $i_1$ and $i_2$. Similarly, let the ground
truth decision be $D_{G}$, we clarify that false negatives may occur, in case
$D_{G} \neq (D_1 = D_2)$.
This may be due to the incorrect (albeit consistent) decisions made by a DNN, or
the analysis errors of our employed XAI algorithms. 
Overall, MT inherently omits considering $D_{G} \neq (D_1 = D_2)$; detecting
such flaws likely requires human annotations, which is highly costly in
real-world settings. On the other hand, as empirically assessed in
\S~\ref{subsec:correctness}, $D$ obtained in this work is accurate.

\subsection{DNN Decision: A Pixel-Based View}
\label{subsec:decision}

We now introduce how a DNN makes decisions. Aligned with previous research, this
paper primarily considers testing DNN image classifiers, and our following
introduction uses image classification as an example accordingly. Many common
DNN tasks root from an accurate image classification (see further discussion in
\S~\ref{sec:discussion}). We first define the \texttt{Empty} and \texttt{Valid}
inputs below.

\vspace{-2pt}
\begin{definition}[\texttt{Empty}]
  \label{def:empty}
  An input is empty if its components are meaningless for humans, e.g., an image
  with random pixel values.
\end{definition}

\vspace{-7pt}
\begin{definition}[\texttt{Valid}]
  \label{def:valid}
  An input is valid if its components are meaningful for humans, e.g., an
  image with human-recognizable objects.
\end{definition}

Given an empty image $\emptyset$, a well-trained DNN $\phi$ will have to
randomly predict a confidence score for each class and the score for class $l$
is $\phi(\emptyset)^l$. A valid input image $i$ can be viewed as introducing the
appearances of its components by changing pixel values over $\emptyset$, namely,
setting $i = \emptyset + \delta$. Accordingly, the output confidence score for
class $l$ is transformed into $\phi(i)^l = \phi(\emptyset)^l + \Delta^l$ given
all these appearances in input. The machine learning community generally views
this procedure as a collaborative game among pixels of
$i$~\cite{lundberg2017unified,ribeiro2016should,bach2015pixel,montavon2019layer,
datta2016algorithmic,sundararajan2017axiomatic,shrikumar2017learning,ancona2019explaining}.
The true contribution of each pixel can be computed via the Shapley
value~\cite{shapley201617} --- a well-established solution in game theory. We
present how to use Shapley value to \texttt{attribute} $\Delta^l$ on $\delta$
below in \Df~\ref{def:attribution}. We then discuss its approximation and
present cost analysis.

\vspace{-2pt}
\begin{definition}[\texttt{Attribution}]
  \label{def:attribution}
  Let each pixel change be $\delta_p$ and $\sum \delta_p = \delta$. Then, an
  \texttt{attribution} of $\Delta^l$ assigns a contribution score $c_p$ to each
  $\delta_p$, such that $\sum c_p = \Delta^l$, where $p$ represents one pixel.
\end{definition}

\parh{From Pixel-Wise Contributions to Decision $D$.}~A pixel $p$
positively supports the DNN prediction for class $l$ if its contribution $c_p >
0$. Therefore, collecting all pixels with positive contributions can help
scoping the decision $D$ upon which DNN $\phi$ relies when processing
$i$ and predicting $l$. Instead of using pixels, however, we
abstract further to group pixels with positive contributions into
visual concepts (e.g., a nose or a wheel) in $i$, and a DNN's predictions can be
decomposed as a voting scheme among visual concepts. Each decision $D$ comprises
all of its visual concepts. We explain how visual concepts are
generated among pixels in \S~\ref{subsec:vc}.

\parh{Approximating Shapley Value in XAI.}~As aforementioned, each pixel in
an image is considered as a player in the collaborative game (i.e., making
a prediction). Let all pixels in an image be $\mathbb{X}$, then calculating
the exact Shapley value requires considering all subset of $\mathbb{X}$ which
results in a computational cost of $2^{|\mathbb{X}|}$ and is infeasible in
practice. Nevertheless, modern
attribution-based XAI~\cite{lundberg2017unified} have enabled practical
approximation of Shapley value. In this research, we use
\dl~\cite{shrikumar2017learning}, a popular XAI tool, to identify pixels $p$ in
an image that positively contribute to the decision of a DNN. 
Though recent works may be able to identify more precise \texttt{attributions}
than \dl, their computation is usually
expensive~\cite{covert2021improving,shrikumar2017learning,lundberg2017unified}.
Also, as noted in \S~\ref{sec:approach}, \dl's potentially imprecise
pixel-level attributions can be cleverly alleviated using our methods.

\begin{figure}[!ht]
  \centering
  \includegraphics[width=0.9\linewidth]{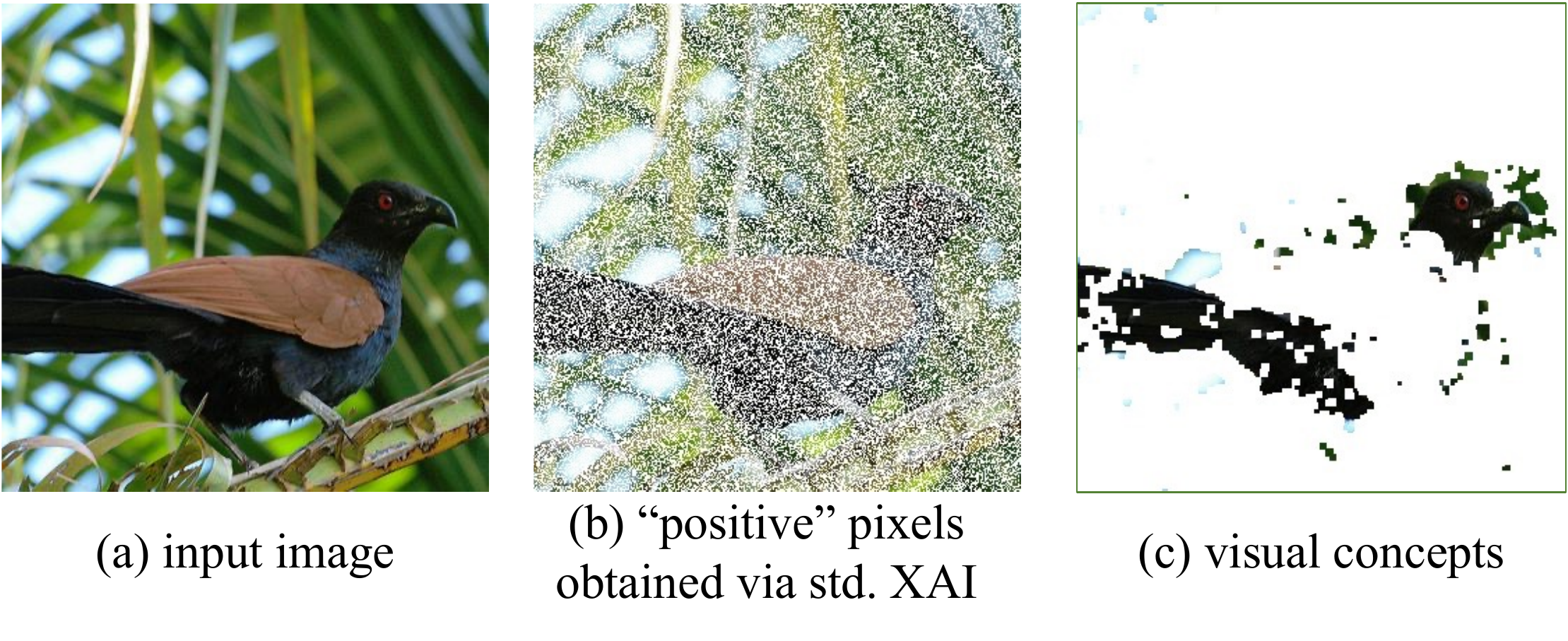}
  \vspace{-12pt}
  \caption{Pixels of positive contributions in standard XAI vs. visual concepts
    converted by our approach.}
  \label{fig:overview}
\end{figure}

\subsection{DNN Decision: A Visual Concept View}
\label{subsec:vc}

It is generally acknowledged that the visual concepts are the basic
perception units captured by humans when perceiving the
world~\cite{kundel1983visual,van2008comparison,mao2015learning}. For instance,
a human may recognize a coucal based on its head and tail, as highlighted in
\F~\hyperref[fig:overview]{2(c)}, rather than those pixels marked in
\F~\hyperref[fig:overview]{2(b)}. We now define the visual concepts:

\begin{definition}[Visual Concept]
  \label{subdef:v-concept}
  A visual concept denotes a semantics meaningful instance on an image, e.g.,
  wheels of a car, or a mark on a car. Generally, each visual concept is a
  connected non-trivial pixel-region and different visual concepts are
  disconnected.
\end{definition}

\parh{Advantage of Visual Concepts.\footnote{The ``visual concept'', which
denotes one region of an input image, is different with what defined in
Activation Atlases~\cite{carter2019exploring} (i.e., one neuron output). }}~As
clarified in \S~\ref{subsec:decision}, we use \dl, a popular XAI tool, to
identify pixels in a DNN input that positively contribute to DNN predictions.
Considering the image (a DNN input) in \F~\hyperref[fig:overview]{2(a)}, we use
\dl\ to identify pixels of positive contributions, as shown in
\F~\hyperref[fig:overview]{2(b)}.
However, ``positive'' pixels span the entire image and their pixel-wise
attribution is \textit{too sensitive} to be used as a testing oracle --- it is
unclear if we have found a ``defect'' where only the contributions of a few pixels
are flipped (e.g., from positively supporting to non-supporting). Moreover, XAI
approaches are not always accurate, as this task is inherently challenging.
In contrast, we abstract pixels into visual concepts for comparison (as in
\F~\hyperref[fig:overview]{2(c)}), which are shown as more robust to potential
inaccuracy of XAI, as trivial pixel-level changes are less likely to differ
visual concepts. 
Moreover, while a single pixel is less interpretable for humans, visual concepts
can explain DNN decisions (e.g., image classification) in a much more
understandable manner; see the following discussion.

\begin{figure}[!ht]
  \centering
  \vspace{-5pt}
  \includegraphics[width=1.0\linewidth]{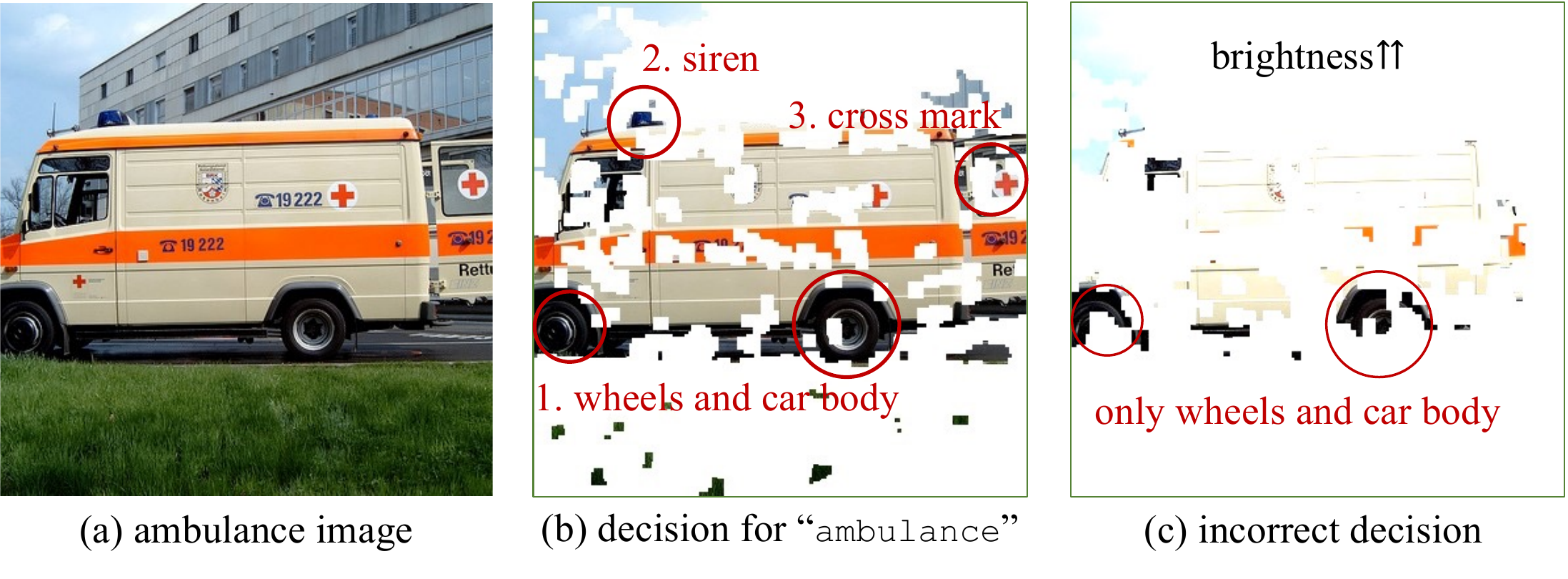}
  \vspace{-20pt}
  \caption{DNN Decision (based on visual concepts) for an ambulance image and
  the one with increased brightness.}
  \vspace{-5pt}
  \label{fig:ambulance}
\end{figure}

\parh{Image Classification based on Visual Concepts.}~Visual concepts 
typically reflect DNN decisions in a human understandable manner. The
prediction ``ambulance'' for \F~\hyperref[fig:ambulance]{3(a)} can be
deconstructed into the visual concepts presented in
\F~\hyperref[fig:ambulance]{3(b)}, where the presence of \texttt{wheels} (visual
concepts) enables the DNN to classify the image as car-like. 
The \texttt{siren} then reduces the possible categories to ``police car'' and
``ambulance.'' The ``ambulance'' is confirmed according to the
\texttt{cross mark}. The \texttt{wheels}, \texttt{siren}, and \texttt{cross
mark} all support the prediction ``ambulance'' in this case. These three visual
concepts form the decision. In other words, the DNN is deemed to
behave incorrectly if certain perturbations result in its decision-making based on distinct
visual concepts. For instance, when the image brightness is
increased (as in \F~\hyperref[fig:ambulance]{3(c)}), the DNN no
longer relies on the \texttt{siren} and \texttt{cross mark} to make decisions.
Despite the DNN still predicts ``ambulance'' (by randomly guessing within
car-like categories), this is deemed a defect in this study.

\parh{Extracting Visual Concepts.}~As noted in \S~\ref{subsec:oracle}, we employ
XAI techniques to mark pixels in a test input that positively contribute to the
DNN's prediction. We further customize and optimize image processing techniques to
construct decision $D$ over those pixels with positive contributions; see
details in \S~\ref{subsec:convert}.

\parh{Comparison with Other Methods.}~Some works obtain DNN decisions
without attribution, but instead use heatmaps whose application scope is
\textit{limited}. For example, the state-of-the-art approach Grad-CAM can fail
if an image has multiple instances of the same
classes~\cite{selvaraju2017grad,chattopadhay2018grad}, which is prevalent in
real images.
In addition, other occlusion-based techniques involve removing/inserting an
object to the input and observing if the prediction changed to decide
decision~\cite{wang2020metamorphic,tian2021extent}. These approaches,
however, require pre-defining and annotating instances in images, which may
impede automated and comprehensive DNN testing.
In contrast, \textit{we generate visual concepts without
human intervention, hence allowing an automated pipeline}.
DNN decisions can also be assessed by observing the DNN internal activities. For
instance, Wang et al.~\cite{wang2020dissector} dissect a DNN as multiple
sub-models and evaluate the internal decision consistency to decide the
prediction correctness. This work is orthogonal to our approach, as we
focus on DNN inputs. Enhancing our approach with DNN internal activities is 
an interesting future work.

\section{Approach Overview}
\label{sec:approach-overview}

By feeding a test image $i$ to an DNN $\phi$, we use $\phi\synbracket{i}.L$ to denote the prediction
(i.e., labels), and $\phi\synbracket{i}.D$ to signify the decision;
$\phi\synbracket{i}.D$ contains a set of visual concepts (fragments) on $i$ that
supports $\phi$ to yield $\phi\synbracket{i}.L$. As introduced in
\S~\ref{subsec:oracle}, this work aims to identify DNN defects that are omitted
by existing MT frameworks using the MR:

\vspace{-5pt}
\begin{equation}
\label{equ:mr}
\mathcal{E}_{l}(\mathcal{M}\synbracket{i}.L, \mathcal{M}\synbracket{i'}.L) \land \mathcal{E}_{d}(\mathcal{M}\synbracket{i}.D, \mathcal{M}\synbracket{i'}.D)
\end{equation}

\parh{Mutating $i$ into $i'$.}~Here, $i'$ is mutated from a test image $i$ using
\mt\ proposed by existing MT-based DNN testing works. For instance, \mt\ can be
pixel-wise transformations or affine transformations. See a full list of \mt\
used in this paper in \S~\ref{sec:implementation}. 

\parh{Checking DNN Predictions.}~$\mathcal{E}_{l}$ is a criterion asserting the
equality of $\phi\synbracket{i}.L$ and $\phi\synbracket{i'}.L$. As introduced in
\S~\ref{subsec:oracle}, this work primarily focuses on DNN image classifiers.
That is, $\phi\synbracket{\cdot}.L$ are image labels, and $\mathcal{E}_{l}$
directly checks the equivalence of two labels.

\begin{figure}[t]
    \centering
    \includegraphics[width=1.0\linewidth]{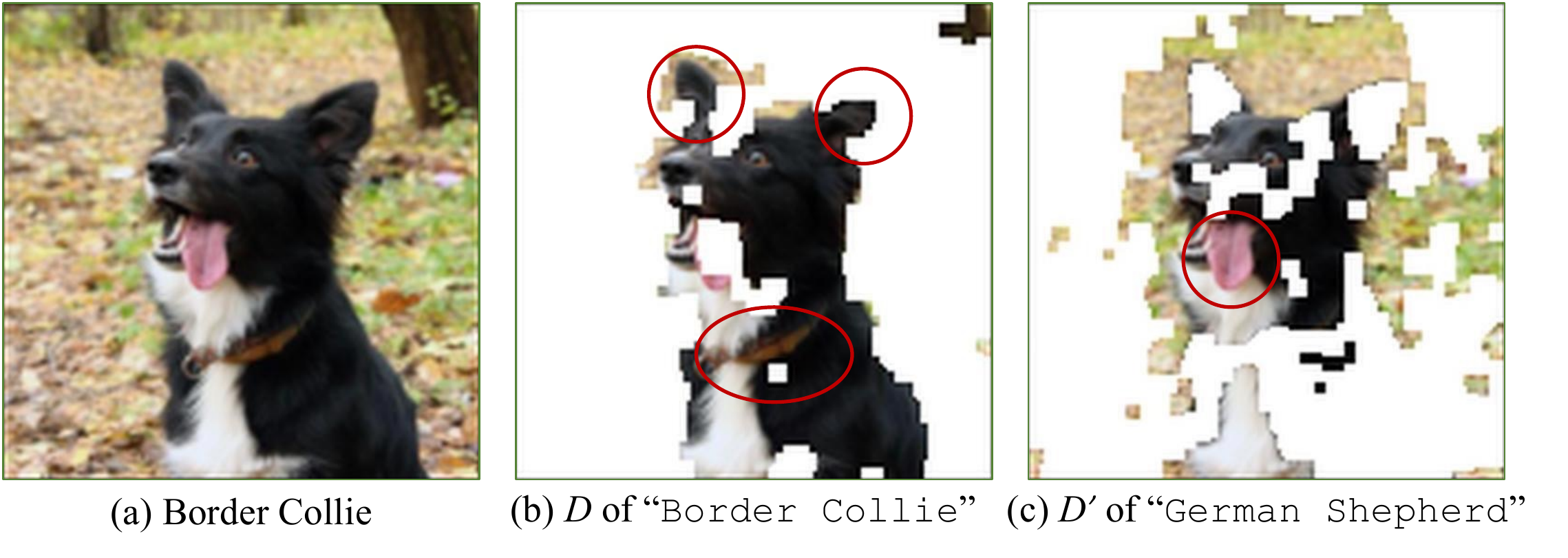}
    \vspace*{-20pt}
    \caption{Inconsistent $D$ and $D'$ when the DNN is making \textit{inconsistent} predictions.
    Decision is \textcolor{pptred}{marked}.}
    \label{fig:case-diff}
\end{figure}

\smallskip
\parhs{Case Study.}~\F~\ref{fig:case-diff} shows a case where the DNN
mis-classifies a ``Border Collie'' image as ``German Shepherd'' when the
original image in \F~\hyperref[fig:case-diff]{4(a)} is slightly blurred. We
present the DNN decision $D$ and $D'$ in \F~\hyperref[fig:case-diff]{4(b)}
and \F~\hyperref[fig:case-diff]{4(c)}, respectively.
As highlighted in \F~\hyperref[fig:case-diff]{4(b)}, the DNN relies on the black
\texttt{ear} and the black-and-white \texttt{neck} to decide the ``Border
Collie.'' In contrast, when \F~\hyperref[fig:case-diff]{4(a)} is blurred as in
\F~\hyperref[fig:case-diff]{4(c)}, the DNN instead focuses on the
\texttt{tongue} and predicts the dog as ``German Shepherd.'' An important
observation, as illustrated in \S~\ref{sec:evaluation}, is that
\textit{$\phi\synbracket{i}.D$ and $\phi\synbracket{i'}.D$ generated using our
decision-based \mr\ are always different when $\phi\synbracket{i}.L \neq
\phi\synbracket{i'}.L$.}

\parh{Checking DNN Decisions.}~$\mathcal{E}_{d}$ is a criterion asserting the
equality of $\phi\synbracket{i}.D$ and $\phi\synbracket{i'}.D$. We introduce an
XAI-based approach to constructing $\phi\synbracket{i}.D$, denoting the decision
of DNN over $i$, in \S~\ref{sec:approach}. Nevertheless, considering each
decision composes a collection of image fragments (each fragment is a
visual concept; see \S~\ref{sec:approach}), asserting the \textit{equality} is
too strict because image fragments could slightly drift without undermining the
DNN overall decisions. 
The computer vision community primarily uses the Intersection over Union (IoU)
metrics to quantify two regions' overlapping. The calculation of IoU will
be given in \S~\ref{subsec:iou}. We deem two decisions violate
$\mathcal{E}_{d}$, in case their overlapping is smaller than a threshold
$T_{iou}$. We empirically decide $T_{iou}$ and present discussions in
\S~\ref{subsec:human}.

\begin{figure}[t]
    \centering
    \includegraphics[width=1.0\linewidth]{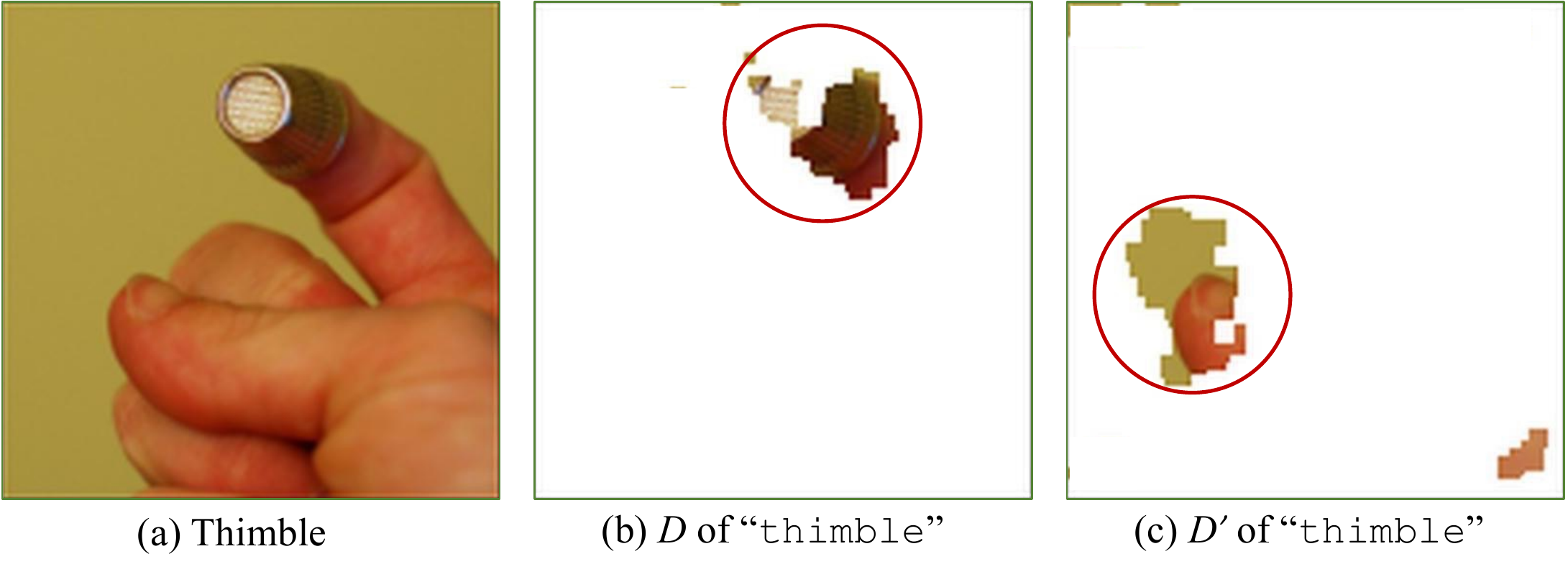}
    \vspace*{-20pt}
    \caption{Inconsistent $D$ and $D'$ when the DNN is making an \textit{identical} prediction.
    Decision is \textcolor{pptred}{marked}.}
    \vspace{-10pt}
    \label{fig:case-same}
\end{figure}

\parhs{Case Study.}~We observe that several $\langle i, i' \rangle$
have zero IoU values for their decisions. Though they lead to the
same prediction, this is likely due to chance. \F~\ref{fig:case-same} shows
a DNN's inconsistent decisions when its predictions are same. 
\F~\hyperref[fig:case-same]{5(a)} is classified as ``thimble'' based
on the highlighted \texttt{thimble} in \F~\hyperref[fig:case-diff]{5(b)}.
However, when the contrast of \F~\hyperref[fig:case-diff]{5(a)} is slightly
lowered, the DNN still predicts ``thimble'' but relies on a new visual concept,
the \texttt{thumb}, as depicted in \F~\hyperref[fig:case-diff]{5(c)}. 
This inconsistency in DNN decisions, which was not previous detected by prediction 
consistency-based MT, suggests that the DNN relies incorrectly on
the thumb to recognize the thimble. We deem this as a hidden defect of this DNN.

\section{Forming and Comparing Decision $D$}
\label{sec:approach}

We clarify the limitation of merely using pixels to form decisions in
\S~\ref{subsec:vc}. Holistically, DNNs are designed to focus on pixel regions
(e.g., using convolutional
kernels~\cite{lecun1989backpropagation,lecun1989handwritten}), and each pixel
should exhibit a closely correlated contribution with its neighbors.
Importantly, pixel-wise contributions can be abstracted into region-wise
\textit{visual concepts}, reflecting a more holistic, robust, and human-readable
view of DNN decisions. We define the decision $D$ as the collection of all
visual concepts in a test input $i$. We have defined \vcc\ in
\S~\ref{subsec:vc}. We present a technical solution to convert pixels (XAI
outputs) to visual concepts in \S~\ref{subsec:convert}, and in
\S~\ref{subsec:iou}, we compare two formed decisions $D$
and $D'$. 

\subsection{Converting Pixels to Visual Concepts}
\label{subsec:convert}

\dl\ marks each pixel in an image with a contribution score. In practice,
because small positive contributions (e.g., $0.001$) are less informative, we
denote pixels with contributions substantially higher than a threshold $T_p$ as
supporting, whereas those with lower or negative contribution scores are
deemed non-supporting. That is, we binarize contributions of a pixel based
on the contribution score \dl\ assigns to each pixel.

\subsubsection{Deciding $T_p$}
\label{subsubsec:threshold}

The above scheme necessitates the use of a threshold $T_p$ to decide the
supporting pixels. A na\"ive approach may be to decide a global threshold.
Nevertheless, pixel-wise contributions vary distinctly between inputs, rendering
a ``global threshold'' less applicable. Consider the case in
\F~\hyperref[fig:ambulance]{3(b)}, where pixels of the \texttt{siren} may
dominate the contributions given a ``police car'' image. However, \texttt{siren}
may contribute less than the \texttt{cross mark} to determine an ``ambulance''
image. The reason is that while \texttt{cross mark} is exclusive to an
ambulance, \texttt{siren} is shared by both classes. 

To overcome this hurdle, our implementation adopts Otsu's
method~\cite{otsu1979threshold} to automatically decide the threshold. In brief,
the Otsu's method seeks the threshold that minimizes intra-class (i.e.,
supporting or non-supporting) variance while simultaneously maximizing the
inter-class (i.e., supporting vs. non-supporting) variance, resulting in a
theoretically optimal threshold with moderate cost.

\subsubsection{Joining and Detaching Supporting Pixels}
\label{subsubsec:join}

Using a threshold $T_p$ decided in \S~\ref{subsubsec:threshold}, we classify
pixels into supporting vs. non-supporting ones. Then, we convert pixels into
visual concepts by joining neighboring pixels that positively support the
prediction and detaching isolated supporting pixels. We first introduce two
elementary image processing operations, \texttt{erosion} and \texttt{dilation},
in this section. For simplicity, we represent the values of supporting and
non-supporting pixels using \texttt{true} and \texttt{false}.

\begin{figure}[!ht]
  \centering
  \includegraphics[width=0.8\linewidth]{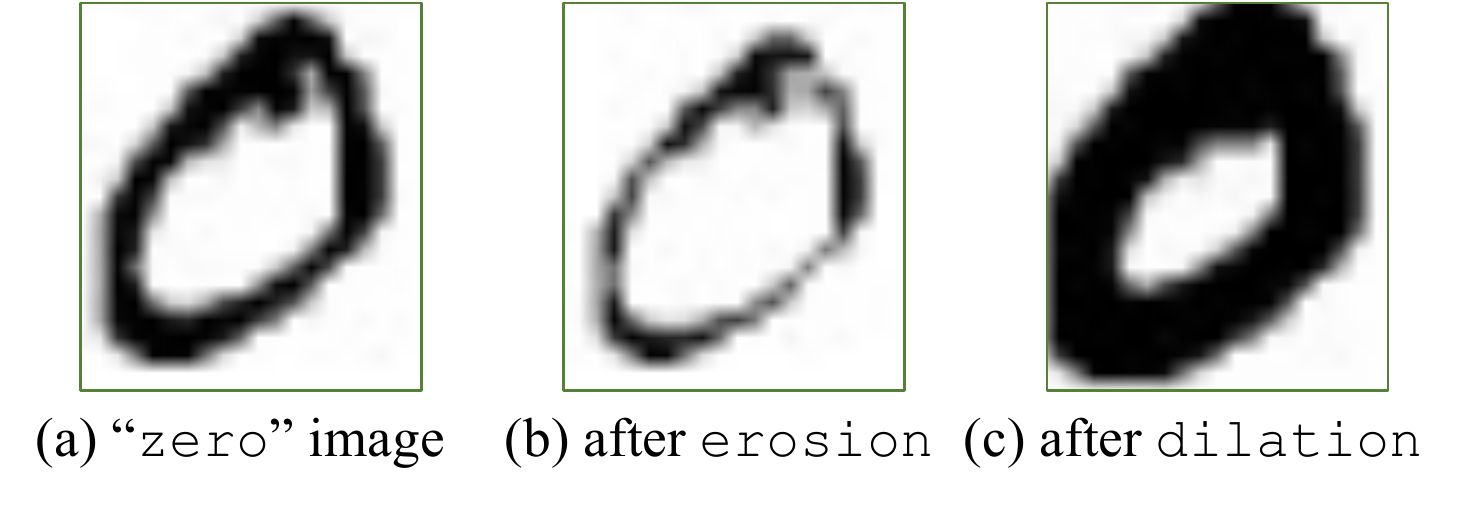}
  \vspace{-15pt}
  \caption{\texttt{Erosion} and \texttt{dilation} transformations. Pixels with
  positive contributions are in black.}
  \label{fig:erosion}
  \vspace{-10pt}
\end{figure}

\parhs{Erosion} operation mimics soil erosion. As in
\F~\hyperref[fig:erosion]{6(b)}, this operation erodes away the boundaries of
positive regions. We implement \texttt{erosion} by using a kernel (similar to a
2D convolution) to slide through the image and retain a positive pixel under a
kernel only if all pixels under the kernel is positive (i.e., their conjunction is
positive). That is, $\texttt{erosion}(p_{k_1, k_2, \dots, k_m}) \leftarrow
\{p_{k_1} \land p_{k_2} \land , \dots, \land p_{k_m}\}^m$.

\parhs{Dilation} is the inverse of erosion. It uses a kernel to slide through
the image but sets a pixel under the kernel as positive as long as any pixel
under the kernel is positive (i.e., their disjunction is positive):
$\texttt{dilation}(p_{k_1, k_2, \dots, k_m}) \leftarrow \{p_{k_1} \lor p_{k_1}
\lor, \dots, \lor p_{k_m}\}^m$. As illustrated in
\F~\hyperref[fig:erosion]{6(c)}, dilation enlarges and sharpens positive
regions, which further increase their visibility.

\begin{figure}[!ht]
  \centering
  \vspace{-10pt}
  \includegraphics[width=0.9\linewidth]{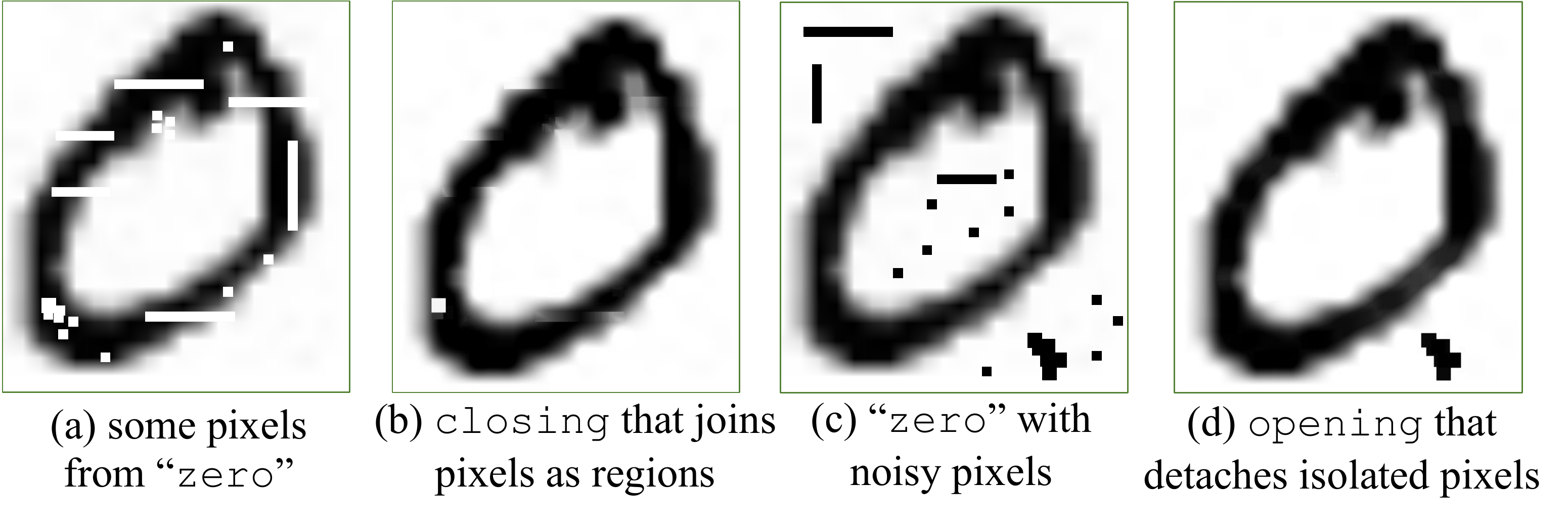}
  \vspace{-15pt}
  \caption{\texttt{Closing} and \texttt{opening} transformations.}
  \label{fig:opening}
  \vspace{-10pt}
\end{figure}

Erosion and dilation are two common operations in image processing. They both
employ a kernel to slide through an input image, resulting in an output image
where the value of each pixel is determined by comparing it to its neighbors in
the input image. Then, depending on the order in which these two basic
operations are used, we obtain the following two advanced operations.

\parhs{Closing} consists of a dilation with a subsequent erosion. That is,
$\texttt{closing}(i) = \texttt{erosion} \circ \texttt{dilation}(i)$. Consider
\F~\hyperref[fig:opening]{7(a)}, the closing operation facilitates fill in small
holes by joining supporting pixels as regions, yielding an image as in
\F~\hyperref[fig:opening]{7(b)}. 

\parhs{Opening} has an erosion followed by a dilation, i.e.,
$\texttt{opening}(i) = \texttt{dilation} \circ \texttt{erosion}(i)$. For
example, \F~\hyperref[fig:opening]{7(c)} will be converted into
\F~\hyperref[fig:opening]{7(d)} after applying \texttt{opening}. This operation
removes small, isolated supporting pixels while retaining the shape and size of
large regions formed by supporting pixels. We employ this operation to remove
isolated supporting pixels which may be likely induced by estimation errors of
XAI.

\parh{From Supporting Pixels to Visual Concepts.}~To convert supporting pixels
into visual concepts, we first apply closing and then opening, i.e.,
$\texttt{to\_concepts}(i) = \texttt{opening} \circ \texttt{closing}(i)$. The
intuition is that, after joining pixels as regions, the remaining isolated
pixels can be regarded as useless (or noisy) for identifying DNN decisions.
The adopted operations are all standard image processing operations.
These operations traverse all pixels in an image in constant time, and can be
computed in parallel. $\texttt{to\_concepts}(\cdot)$ can be
implemented by replacing the kernel in a 2D convolutional layer, which slides
through the image and is multiplied with its covered region, with a sequence
of conjunction and disjunction operations detailed in this section. 2D
convolutional layers are common in image-processing DNNs. Computation cost
incurred at this step is thus \textit{equivalent to adding one DNN layer.}
Moreover, the kernel size in \texttt{closing} and \texttt{opening} decides the
effectiveness of extracting visual concepts: a larger kernel is more abstract
and thus more robust to potential XAI errors, but less precise. We set the
kernel size of \texttt{closing} and \texttt{opening} to $5$ and $3$, which are
two most widely used kernel sizes in convolutional neuron networks that
characterize the ``spotting region'' of common DNNs.

\subsection{Measuring Inconsistent $D$ using IoU}
\label{subsec:iou}

Our observation shows that hidden DNN defects unveiled using our MR in
\E~\ref{equ:mr} belong to either of the following categories or their
composition.

\noindent 1) \underline{Missing visual concepts} occurs when the tested DNN
relies on a subset of the original visual concepts for prediction. As in
\F~\ref{fig:ambulance}, suppose the tested DNN neglects the \texttt{cross mark}
under certain perturbations, it has to output ``police car'' or ``ambulance'' by
chance according to our analysis. By checking ``missing visual concepts,'' we
unveil DNN defects even if the prediction is still ``ambulance.''
  
\noindent 2) \underline{New visual concepts} occurs if the tested DNN uses
irrelevant visual concepts in the image to make decisions. Recall our
motivating example in \F~\ref{fig:motivation}, where the DNN relies on a flower
in \F~\hyperref[fig:motivation]{1(d)} as a new visual concept for
prediction. Such incorrect decision can also be uncovered by checking the
decision consistency.

\begin{figure}[!htp]
  \vspace*{-10pt}
  \centering \includegraphics[width=1.0\linewidth]{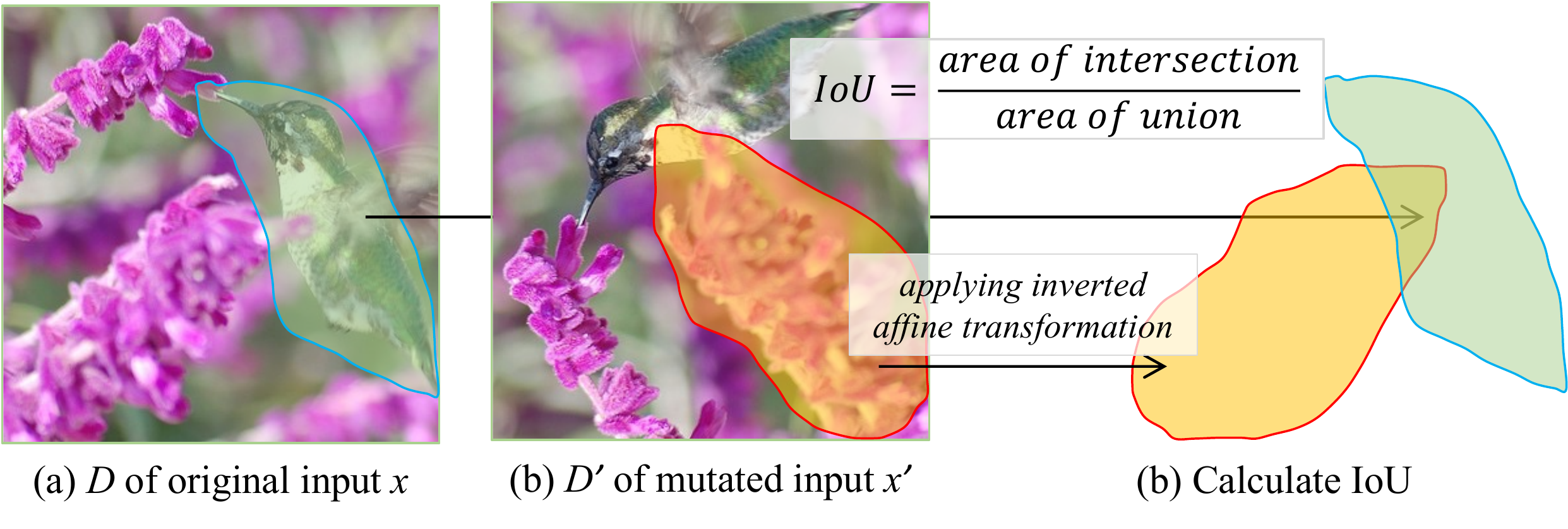}
  \vspace*{-15pt}
  \caption{IoU calculation. For this case,
  since $x'$ is generated by using affine transformation over $x$, the IoU is
  calculated after inverting this affine transformation on $D'$.}
  \vspace*{-3pt}
  \label{fig:iou}
\end{figure}

Overall, the above cases should \textit{not} happen since we mutate test inputs
using \textit{semantics-preserving} \mt. See \mt\ adopted in
this research in \S~\ref{sec:implementation}.  We use IoU to measure the
overlapping of decisions $D$ and $D'$ captured in a test input and its mutated
input. The calculation of IoU is illustrated in \F~\ref{fig:iou}. Overall, each
IoU value is a number from 0 to 1 that quantifies the degree of overlapping
between two regions. 

\parh{Clarification on MR.}~Careful readers may wonder if
the above oracle is a bit too strong, because adding/missing a non-important visual concept
may not affect DNN predictions from human perspectives, e.g., missing one \textit{cross mark} may not
alter predicting ``ambulance'' for \F~\ref{fig:ambulance}. Nevertheless, besides
classification, features (i.e., the outputs of intermediate layers)
extracted from classifiers typically facilitate (security-sensitive) downstream
tasks like object tracking and auto-driving, where adding/missing visual concepts can be more
serious and likely lead to severe outcomes~\cite{he2017mask,girshick2015fast,ren2015faster}.
Moreover, to take into account cases where inconsistent visual concepts are
valid, we employ human evaluations to decide a threshold (see \textbf{RQ2} in
\S~\ref{subsec:human}) to better decide visual concept ``inconsistency.''

\begin{table}[h]
    \centering
    \caption{Metamorphic transformations \mt\ adopted.}
    \label{tab:mt}
    \vspace{-10pt}
    \resizebox{0.9\linewidth}{!}{
    \begin{tabular}{
        @{\hspace{1pt}}l@{\hspace{2pt}}|
        Sl|c}
    \hline
    \textbf{\mt} & \textbf{Original} $x$ $\Rightarrow$ \textbf{Mutated Image} $x'$ & Used by\\
    \hline
    \shortstack{Pixel\\Level} &
            \makecell{\rowincludegraphics[scale=0.08]{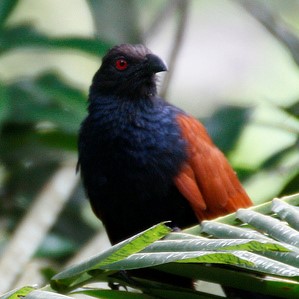}\\{\footnotesize $x$}}
            $\Rightarrow$
            \makecell{\rowincludegraphics[scale=0.08]{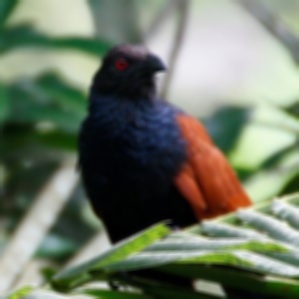}\\{\footnotesize blur}},
            \makecell{\rowincludegraphics[scale=0.08]{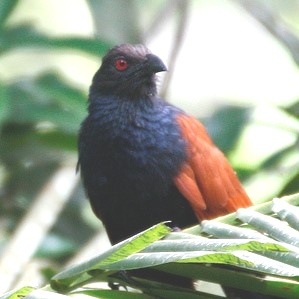}\\{\footnotesize brightness}}, 
            \makecell{\rowincludegraphics[scale=0.08]{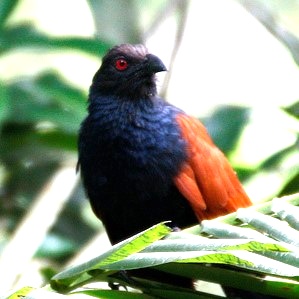}\\{\footnotesize contrast}}
    & \shortstack{\cite{pei2017deepxplore},\cite{tian2018deeptest},\\
    \cite{xie2018coverage},\cite{demir2019deepsmartfuzzer}} \\
    \hline
    \shortstack{Affine\\Type} &
            \makecell{\rowincludegraphics[scale=0.08]{fig/mutate/org.jpg}\\{\footnotesize $x$}}
            $\Rightarrow$
            \makecell{\rowincludegraphics[scale=0.08]{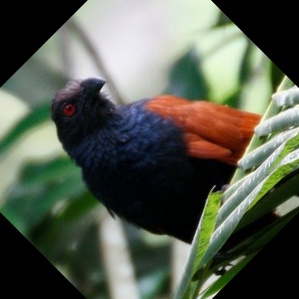}\\{\footnotesize rotation}},
            \makecell{\rowincludegraphics[scale=0.08]{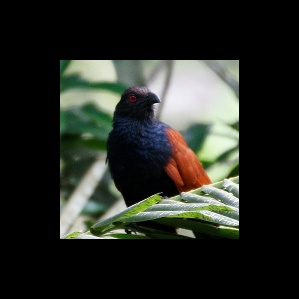}\\{\footnotesize scale}}, 
            \makecell{\rowincludegraphics[scale=0.08]{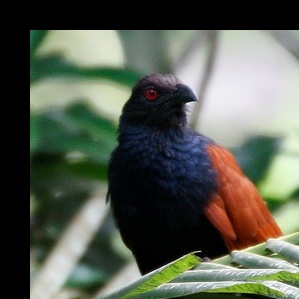}\\{\footnotesize translation}}, 
            \makecell{\rowincludegraphics[scale=0.08]{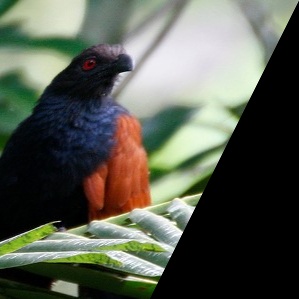}\\{\footnotesize shear}}
    & \shortstack{\cite{pei2017deepxplore},\cite{tian2018deeptest},\\
    \cite{xie2018coverage},\cite{demir2019deepsmartfuzzer}} \\
    \hline
    \shortstack{Weather\\Filter} &
    \makecell{\rowincludegraphics[scale=0.08]{fig/mutate/org.jpg}\\{\footnotesize $x$}}
            $\Rightarrow$
            \makecell{\rowincludegraphics[scale=0.08]{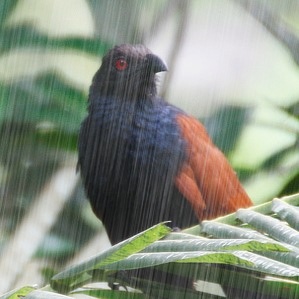}\\{\footnotesize rainy}},
            \makecell{\rowincludegraphics[scale=0.08]{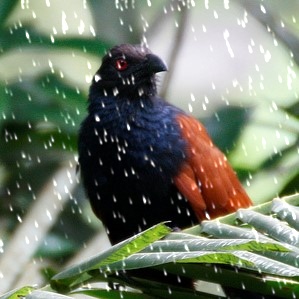}\\{\footnotesize snowy}}, 
            \makecell{\rowincludegraphics[scale=0.08]{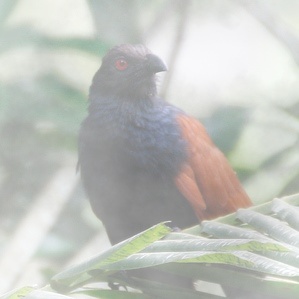}\\{\footnotesize foggy}},
            \makecell{\rowincludegraphics[scale=0.08]{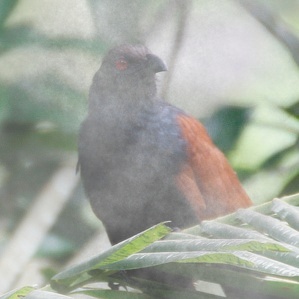}\\{\footnotesize cloudy}}
    & \cite{tian2018deeptest} \\
    \hline
    \shortstack{Style\\Transfer} &
    \makecell{\rowincludegraphics[scale=0.08]{fig/mutate/org.jpg}\\{\footnotesize $x$}}
            $\Rightarrow$
            \makecell{\rowincludegraphics[scale=0.08]{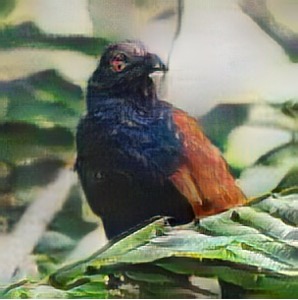}\\{\footnotesize style 1}},
            \makecell{\rowincludegraphics[scale=0.08]{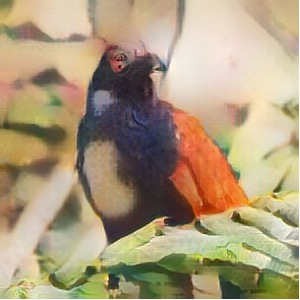}\\{\footnotesize style 2}}, 
            \makecell{\rowincludegraphics[scale=0.08]{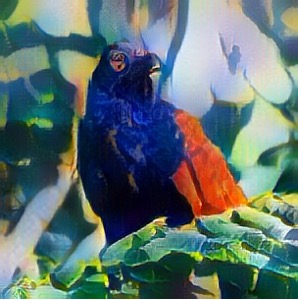}\\{\footnotesize style 3}},
            \makecell{\rowincludegraphics[scale=0.08]{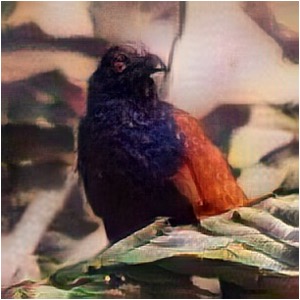}\\{\footnotesize style 4}}
    & \cite{zhang2018deeproad} \\
    \hline
    \shortstack{Adv.\\Perturb.} & 
    \makecell{\rowincludegraphics[scale=0.08]{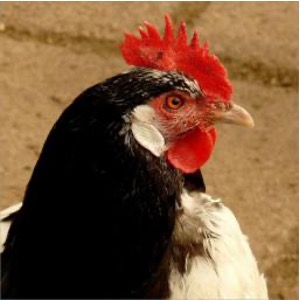}\\{\footnotesize $x$}}
            $+=$
            \makecell{\rowincludegraphics[scale=0.08]{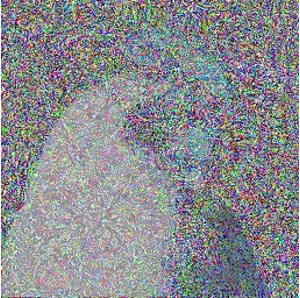}\\{\footnotesize PGD}},
            \makecell{\rowincludegraphics[scale=0.08]{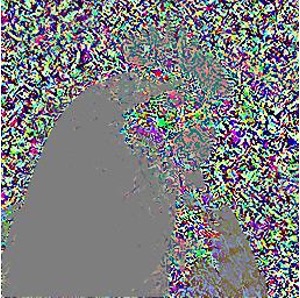}\\{\footnotesize C/W}}, 
            \makecell{\rowincludegraphics[scale=0.08]{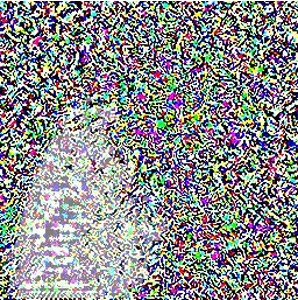}\\{\footnotesize FGSM}},
            \makecell{\rowincludegraphics[scale=0.08]{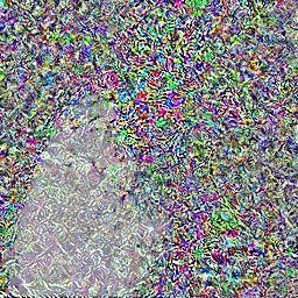}\\{\footnotesize BIM}}
    & \shortstack{\cite{goodfellow2014explaining},\cite{madry2017towards},\\
    \cite{carlini2017towards},\cite{kurakin2016adversarial}} \\
    \hline
    \end{tabular}
    }
\vspace{-10pt}
\end{table}

\section{Implementation \& Setup}
\label{sec:implementation}

The entire codebase of this research, including scripts to form decision-based
\mr\ and extend prior \mt, has about 1500 lines of Python code. 
We use \dl, a popular XAI framework, to flag pixel-wise
contributions. We reuse and extend \mt\ implementation of existing MT-based DNN
testing frameworks. Most existing MT-based DNN testing works
test image classification models by mutating images. \mt\
in prior works are essentially semantics-preserving image mutations, as reviewed below.

\parh{Pixel-Level Mutations.}~This scheme contains multiple instances. 
By adding a same constant to all pixels of an image, we can change
image brightness. Similarly, multiplying all pixels with a same constant can
change image contrast. Blurring is implemented by multiplying the image
with a sliding kernel. These three mutations are adopted to assess the
robustness of DNNs~\cite{pei2017deepxplore,tian2018deeptest,xie2018coverage,
odena2018tensorfuzz,demir2019deepsmartfuzzer}.

\parh{Affine Transformations.}~This scheme mutates objects in images by
applying invertible transformations such as translating, rotating, scaling, and
shearing. These transformations preserve the collinearity of objects, thereby
retaining correct semantics of objects after mutation. These methods are
used to diverse
images~\cite{tian2018deeptest,demir2019deepsmartfuzzer,xie2018coverage}.

\parhs{Clarification.}~Recall that our \mr\ checks the consistency of decision
$D$. Nevertheless, affine transformations, by mutating objects in an
image, unavoidably change the localization of visual concepts. Thus, we calculate
the IoU after inverting the applied affine transformation on
generated visual concepts. For instance, if the applied affine transformations
left rotate 90 degrees, we measure the IoU after first right rotating the
visual concepts by 90 degrees.

\parh{Weather Filters.}~To simulate various weather conditions in real
scenarios, weather filters, including snowy, foggy, rainy, and cloudy, are
widely adopted to mutate test inputs of DNNs. We adopt weather filters
proposed in~\cite{tian2018deeptest,imgaug} for mutation.

\parh{Style Transfer.}~Style transfer was first proposed
by~\cite{zhang2018deeproad}, which is specifically launched to transfer (severe)
weathers from a source image to a target image. The generated weather conditions
in the target are usually in a higher quality than using weather filters. This
approach, however, is limited to driving scenes. We extend it to
general style-transfer such that arbitrary real images can be mutated.
Style transfer primarily mutates image colors and retain the semantics.
Our mutations include 72,521 styles; see our code~\cite{snapshot}.

\parhs{Remarks.}~The weather filters and style transfer are only applied on
CIFAR10 and ImageNet which consist of real images. Images in MNIST, a small
synthetic dataset, are handwritten digits where no ``weather'' exists. It is
also infeasible to change the color scheme via style transfer, given images in
MNIST are black-and-white without the color channel (e.g., the RGB channel in
real images).

\parh{Adversarial Perturbations.}~Guided by DNN gradients, adversarial
perturbations generate adversarial examples (AEs) to fool DNNs. AEs are usually
less notable or visually identical to source images. As shown in
\T~\ref{tab:mt}, adversarial perturbations often highlight key objects in an
image. We employ four popular algorithms, FGSM~\cite{goodfellow2014explaining},
PGD~\cite{madry2017towards}, C/W~\cite{carlini2017towards} and
BIM~\cite{kurakin2016adversarial}, to generate AEs.

The above mutations are frequently used in previous works: with
carefully chosen parameters (i.e., to what extent the mutation is applied),
they generally retain the semantics (visual consistency) in the mutated
images. Therefore, a functional DNN should preserve its decisions under
these mutations. In experiments, we ship \mt\ configurations from previous
works~\cite{pei2017deepxplore,xie2018coverage,tian2018deeptest,
demir2019deepsmartfuzzer,hu2022if} and only mutate a seed (a real image)
once; see our implementation in~\cite{snapshot}.

\parh{Preparing Datasets $\hat{I}_{=}$ and $\hat{I}_{\neq}$.}~We prepare an
image set $I$ by randomly selecting images from a dataset listed in
\T~\ref{tab:model}. These three datasets are the most widely-used datasets
for DNN testing. For each $i \in I$, we mutate it with different $\texttt{t}
\in MR_{t}$, as presented in \T~\ref{tab:mt}, and produce $i_\texttt{t} =
\{\texttt{t}(i) | \texttt{t} \in MR_{t}\}$.\footnote{Note that a \mt\
$\texttt{t}$ may be used for multiple times, each with a randomly selected
configuration, e.g., rotation degree or brightness level.} Then, for each $i$
and its mutated $i' \in i_\texttt{t}$, we construct a set $\hat{I}$ with image
pairs $\langle i, i' \rangle$, namely, $\hat{I} = \{ \langle i, i' \rangle | i
\in I, i' \in i_\texttt{t} \}$. We further divide $\hat{I}$ into two
collections: 1) $\hat{I}_{=}$ where the DNN $\phi$ has the same prediction label
for $i$ and $i'$, i.e., $\hat{I}_{=} = \{ \langle i, i' \rangle |
\phi\synbracket{i}.L = \phi\synbracket{i'}.L, \langle i, i' \rangle \in \hat{I}
\}$, and 2) $\hat{I}_{\neq}$ where the DNN $\phi$ yields inconsistent labels for
$i$ and $i'$, i.e., $\hat{I}_{\neq} = \{ \langle i, i' \rangle |
\phi\synbracket{i}.L \neq \phi\synbracket{i'}.L, \langle i, i' \rangle \in
\hat{I} \}$.
For each tested DNN, we keep generating $\hat{i}$ until both $\hat{I}_{=}$ and
$\hat{I}_{\neq}$ has $10,000$ pairs. 

\smallskip
\parh{Preparing Tested DNNs.}~We list all DNN models tested in this paper in
\T~\ref{tab:model}. Note that for the first three models, we use two instances
trained over two different datasets, ImageNet and
CIFAR10. All ImageNet-trained models are officially
provided by PyTorch and other models are well-trained (over 94\% test accuracy).
They are commonly used in daily DNN tasks and testings.

\begin{table}[t]
    \caption{Evaluated DNN models.}
     \vspace{-10pt}
    \label{tab:model}
    \centering
  \resizebox{0.9\linewidth}{!}{
    \begin{tabular}{l|c|c}
      \hline
       Model         & Dataset & Remark \\
      \hline
      ResNet50~\cite{he2016resnet}     & ImageNet~\cite{imagenet}/CIFAR10~\cite{krizhevsky2009cifar}  & Non-sequential structure \\
      \hline
      VGG16~\cite{simonyan2014very}   & ImageNet/CIFAR10  & Sequential structure \\ 
      \hline
      MobileNet-V2~\cite{howard2017mobilenets}  & ImageNet/CIFAR10  & Mobile devices \\ 
      \hline
      DenseNet121~\cite{huang2017densely}  & ImageNet  &  Extremely deep model \\ 
      \hline
      Inception-V3~\cite{szegedy2016rethinking}  & ImageNet  &  Feature representation \\ 
      \hline
      LeNet1~\cite{lecun1989backpropagation}  & MNIST~\cite{deng2012mnist}  &  Black-white images \\ 
      \hline
      LeNet5~\cite{lecun1998gradient}  & MNIST &  Black-white images \\ 
      \hline
    \end{tabular}
    }
    \vspace{-5pt}
\end{table}

\section{Evaluation}
\label{sec:evaluation}

We primarily study the following research questions.

\noindent \textbf{RQ1}: Is the identified decision $D$ in each DNN input correct?
We answer this question in \S~\ref{subsec:correctness}.

\noindent \textbf{RQ2}: Are inconsistency of decisions $D_1 \neq D_2$ truly reflect
DNN defects? We launch large-scale human evaluation to explore this question
from different aspects in \S~\ref{subsec:human}.

\noindent \textbf{RQ3}: How many and how frequent hidden defects are overlooked by
existing MT-based DNN testing? What are the characteristics of these hidden
defects? We answer this question in \S~\ref{subsec:empirical}.

\begin{table}[t]
  \caption{Correctness of identified decisions. We report the percentage
  of unchanged outputs (left) when only keeping the decisions, and the
  percentage of changed outputs (right) when masking decisions in inputs.
  }
   \vspace{-5pt}
  \label{tab:corret}
  \centering
\resizebox{0.95\linewidth}{!}{
  \begin{tabular}{c|c|c|c|c}
    \hline
    \shortstack{ResNet50\\ImageNet} & \shortstack{VGG16\\ImageNet} & \shortstack{MobileNetV2\\ImageNet} & \shortstack{DenseNet121\\ImageNet} & \shortstack{Inception-V3\\ImageNet} \\
    \hline
    96.0\%, 100\% & 96.8\%, 100\% & 96.5\%, 100\% & 96.9\%, 100\% & 95.8\%, 100\% \\
    \hline
    \shortstack{ResNet50\\CIFAR10} & \shortstack{VGG16\\CIFAR10} & \shortstack{MobileNetV2\\CIFAR10} & \shortstack{LeNet1\\MNIST} & \shortstack{LeNet5\\MNIST} \\
    \hline
    97.6\%, 93.8\% & 95.8\%, 92.6\%,  & 96.8\%, 92.7\% & 96.2\%, 91.8\% & 96.9\%, 93.5\% \\
    \hline
  \end{tabular}
  }
  \vspace{-10pt}
\end{table}

\subsection{RQ1: Correctness of Decisions}
\label{subsec:correctness}

It is challenging to assess the correctness of obtained decisions: as
clarified in \S~\ref{subsec:oracle}, we lack the ground truth decisions $D_{G}$
for diverse real-world images. Furthermore, given that DNNs are designed to
discover (subtle) patterns inherent in the data, it is likely that they make
decisions using visual concepts that are diametrically different from human
judgments, but still valid.

To measure the correctness of captured decisions, we aim to construct
\textit{contradictory facts} to the DNN to assess the accuracy of
decisions. Our intuition is that if and only if the identified visual
concepts correctly constitute the decisions of the DNN, the DNN prediction
should \textit{not} be changed when only these visual concepts are extracted to form DNN
inputs. Accordingly, if the identified visual concepts in an input are masked,
the DNN prediction should change (as it loses the decisions).

For each tested DNN, \textit{the above schemes are performed for the original inputs
and the mutated input pairs from both $\hat{I}_{=}$ and $\hat{I}_{\neq}$}. Results
are reported in \T~\ref{tab:corret}. We also use the above masking schemes
directly toward pixels identified by conventional XAI techniques, including
raw outputs of \dl. For both two masking schemes, they only have
around 50\% of outputs unchanged/changed.
In contrast, the converted visual concepts are more reliable, as it better
correlates with our expectation: when identified visual concepts are masked,
over $1 - \frac{1}{C}$ of outputs are changed (100\% in ImageNet cases), where
$C$ is the \#classes. Since the DNN has $\frac{1}{C}$ chance to guess the
output, our identified decisions are precise. Similarly, when inputs only
contain the identified decisions, all models have over $95\%$ outputs
unchanged. Most identified visual concepts are small fragments in images (they
may be ``invalid'' as an input image). Thus, it is reasonable that the results
are slightly lower than 100\%. Overall, we interpret that decisions
extracted by our technique represent the true decisions of various DNNs.

\begin{tcolorbox}[size=small]
  \textbf{Answer to RQ1}: With experiments based on contradictory facts, we
  illustrate the accuracy of extracted decisions. We also find that the
  decision, denoting regions on images, are more reliable than
  merely the pixels extracted by XAI.
\end{tcolorbox}

\subsection{RQ2: Human Evaluation}
\label{subsec:human}

We now analyze how IoU values, which are derived by comparing the visual
concepts of an image pair $\langle i, i' \rangle$, of $\hat{I}_{=}$ and
$\hat{I}_{\neq}$, are distributed. 
We first report that for all image pairs in $\hat{I}_{\neq}$, the maximal IoU
value $v < 0.8$, indicating that visual concepts among $i$ and $i'$ are
\textit{always different} whenever the DNN yields different labels. 
Moreover, we report that the IoU value $v$ calculated over images in
$\hat{I}_{=}$ ranges from $0$ to $1$. Obviously, an IoU value $v = 1$ indicates
that the decisions of $i$ and $i'$ are exactly the same, falling into the
\CircleOne\ case in \T~\ref{tab:decision} where $L_1 = L_2$ and $D_1 = D_2$. An
IoU $v = 0$, in contrast, illustrates that the decisions in $i$ and $i'$
are deemed as different. For such cases, it is clear that the DNN makes an
incorrect prediction, falling into scenario \CircleThree\ of 
\T~\ref{tab:decision} where $L_1 = L_2$ and $D_1 \neq D_2$. 

When the IoU value $0 < v < 1$, it is unclear if DNN makes incorrect
predictions. A simple method is to consider two decisions distinct as long as $v
< 1$. However, our manual analysis shows that for $\hat{I}_{=}$, $v < 1$ does
not necessarily imply that the DNN makes different decisions over $i$ and $i'$.
For instance, the non-overlapping may be caused by different fragments on grass
and clouds and other large objects in $i$. Despite the existence of
non-overlapping visual concepts, they all represent the same object. Similarly,
a relatively higher (near to $1$) IoU value does not always indicate no defect,
as some \textit{small yet critical} visual concepts may cover only a small
portion of areas among all identified ones (e.g., the cross mark in an
ambulance; see \F~\ref{fig:ambulance}).

We conduct human evaluations to explore to what extent the IoU value can reflect
the inconsistency of two decisions $D$ and $D'$. We form a group of $25$
participants to answer a total of $10,000$ questions. Given images from CIFAR10
and MNIST have relatively lower resolution, we only use images from ImageNet ---
the de facto real-life dataset which has the largest scale among all evaluated
datasets. All evaluations are performed on the Amazon Mechanical Turk platform.
We explain the setup and give quantitative results in
\S~\ref{subsubsec:pg-student}. We present qualitative feedback in
\S~\ref{subsubsec:feedback}.

\begin{figure}[!ht]
  \centering
  \includegraphics[width=1.01\linewidth]{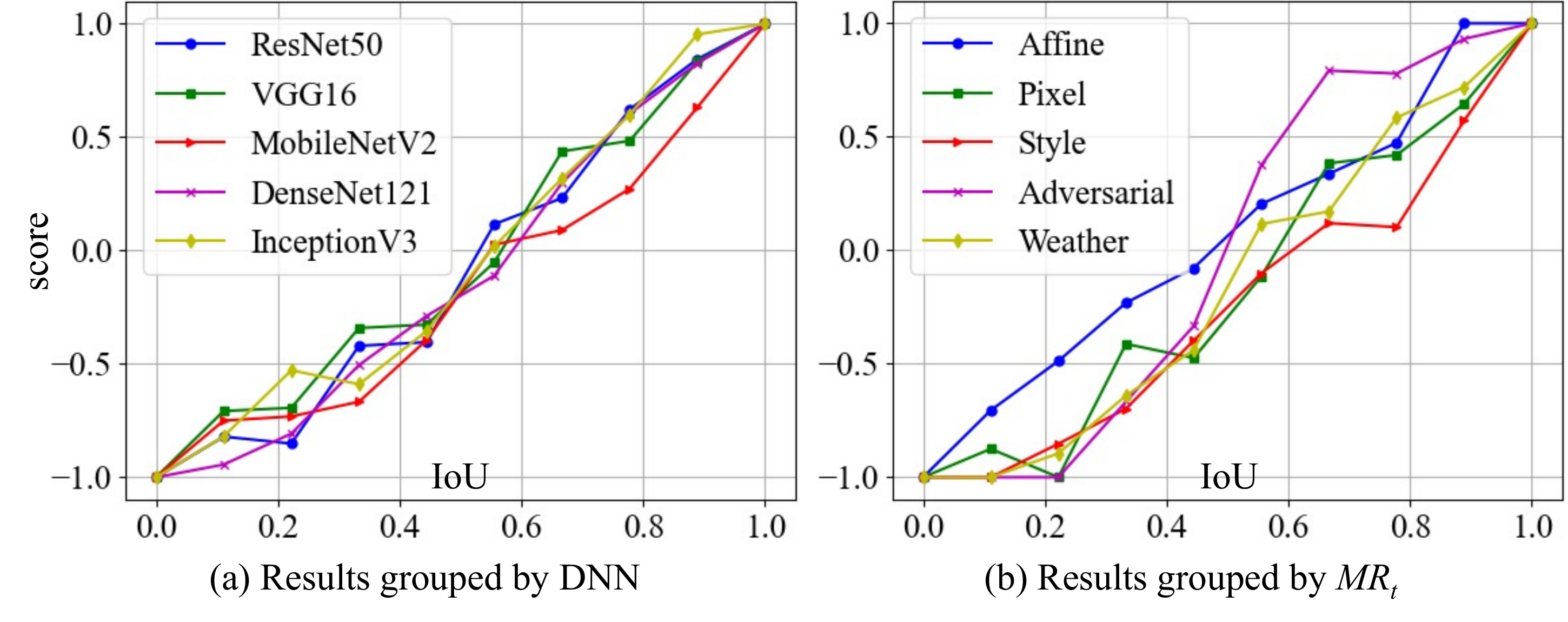}
  \vspace{-25pt}
  \caption{Human evaluation results of \S~\ref{subsubsec:pg-student}.}
  \label{fig:human1}
  \vspace{-5pt}
\end{figure}

\begin{figure}[!ht]
  \vspace{-5pt}
  \centering
  \includegraphics[width=1.01\linewidth]{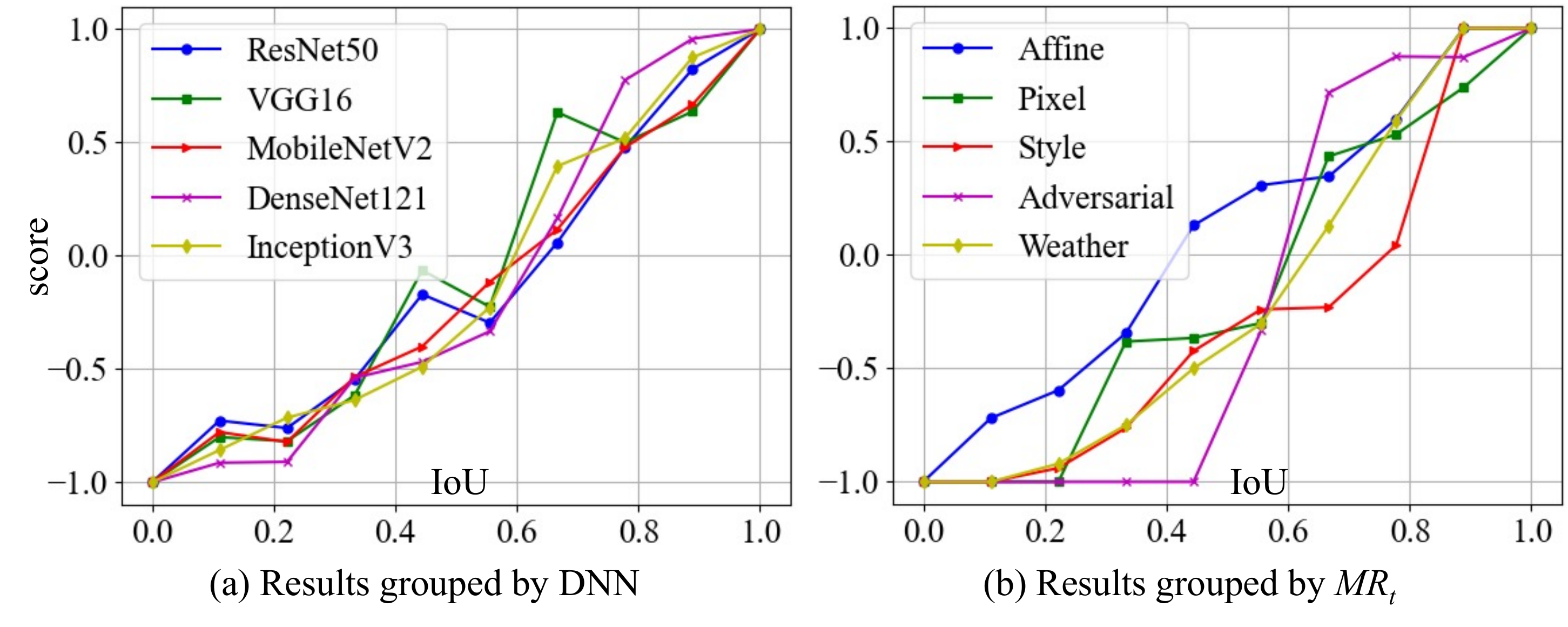}
  \vspace{-20pt}
  \caption{Human evaluation results of \S~\ref{subsubsec:non-expert}.}
  \label{fig:human2}
  \vspace{-10pt}
\end{figure}

\subsubsection{Are Decisions Same?}
\label{subsubsec:pg-student}

Since understanding the visual concept and DNN decisions may be obscure for
laymen, we invite $15$ Ph.D. students with experience in computer
vision and DNN related projects to participate this evaluation. Prior to the
experiment, we teach a student what a visual concept is through examples and
explain how to distinguish consistent/inconsistent DNN decisions, until the
student is confident that he/she understands the prior knowledge. The teaching
takes in average $30$ minutes for each participant; we provide full
materials used at this step to benefit future research on our released
artifact~\cite{snapshot}.

We use five popular and representative DNNs trained on ImageNet, as listed in
\T~\ref{tab:model}. For each model, we randomly select $500$ pairs of images,
which are evenly generated by each type of \mt\ we introduced in
\S~\ref{sec:implementation}, from $\hat{I}_{=}$ and collect a set of total
$2,500$ pairs. We then repeat each pair three times so that three participants
can analyze each pair of decisions. The totaling 7,500 images are evenly
assigned to $15$ participants. Each participant may halt the human evaluation
anytime to avoid fatigue. On average, each participant requires $2.5$ hours to
finish the entire evaluation.

For each question, we present the student three images: the original image $i$
and two images displaying only the decisions of $i$ and $\hat{i}$. Then, we ask
whether the two decisions are equivalent (i.e., they rely on the same set of
visual concepts). Students may select ``\texttt{Yes}'', ``\texttt{No}'', or
``\texttt{Not sure}'', which count 1, -1, 0, respectively.
As aforementioned, since DNNs are designed to explore (subtle) patterns in data,
it can be difficult for humans to deduce the image content given only the visual
concepts captured by DNNs. By providing the original image $i$, students grasp
the ``whole picture'' in the image, and we find this helps them compare
decisions in $i$ and $\hat{i}$. Otherwise, they may just compare the pixel-wise
difference, if they have no idea about what is displayed in the image. 

Moreover, we find that students may fail to truly understand the requirements of
this experiment, resulting in potentially invalid findings. As a common
practice, we prepare and insert a number of sanity-check (SC) questions randomly
into the raised questions without prior notice. Students must correctly answer
SCs in order to justify the reliability of their answers. Specifically, in some
questions, we present two identical decisions (i.e., extracted from the
same image), for which participants should answer ``yes.'' We also reorder the
two images of decisions in some questions to generate duplicated questions
--- reordering makes students less likely to notice that they have already answered
this question; for duplicated questions, student should answer consistently
with the original questions. We discard answers from students who
correctly answer $< 95\%$ SC questions. \textit{All participants have passed the
sanity check.}

To analyze the results, we group questions by 1) model and 2) the type of \mt.
We show how the scores of questions change with IoU values in
\F~\ref{fig:human1}. It is evident that each question's score has a strong
positive correlation with the IoU value. When the decisions of $\langle i, i'
\rangle$ have a small IoU value ($< 0.2$), nearly all participants agree that
they denote \textit{different} decisions. Similarly, most participants regard
two decisions as the same if they have a sufficiently higher IoU value ($> 0.8$).
From \F~\ref{fig:human1}, we notice that in all cases (i.e., grouped by $5$
models, or $5$ types of \mt), scores of $\langle i, i' \rangle$ pairs are less
than 0 when the corresponding IoU values are smaller than $0.5$. Moreover, the
score is around $-0.75$, when the IoU values are smaller than $0.2$. This
indicates the majority of participants agree that the two decisions are
different. As a practical setup, we recommend IoU values in the vicinity of $0.2$.

\subsubsection{How Apparent the Decisions Change?}
\label{subsubsec:non-expert}

To better justify our findings, we further conduct an experiment to evaluate
how apparent the decision changes are via assessing whether they are
human-perceptible \textit{at the first glance}. 
The intuition is that, if the difference can be noticed at the first glance,
it is likely induced by distinct visual concepts and thus indicates a DNN defect.
This experiment follows mostly the same setup as in
\S~\ref{subsubsec:pg-student} but only shows the decisions of $i$ and $i'$
\textit{two seconds}. Moreover, to make the results more general (students with
different backgrounds may have distinct preference; see
\S~\ref{subsubsec:feedback}), we invite another 10 Ph.D. and masters students
with \textit{various backgrounds} to evaluate total $2,500$ pairs of decisions
from $\hat{I}_{=}$; see detailed setups in~\cite{snapshot}. We present results
in \F~\ref{fig:human2}. Compared to \F~\ref{fig:human1}, results in
\F~\ref{fig:human2} has a slightly higher fluctuation, which may be due to the
participants' small time window (2 seconds). Nevertheless, we observe that
trends in \F~\ref{fig:human1} and \F~\ref{fig:human2} are highly consistent,
indicating that the decisions are \textit{highly recognizable} and of
\textit{high quality}.

\subsubsection{Agreements of Participants}
\label{subsubsec:kappa}

\begin{figure}[!ht]
  \centering
  \vspace{-5pt}
  \includegraphics[width=1.01\linewidth]{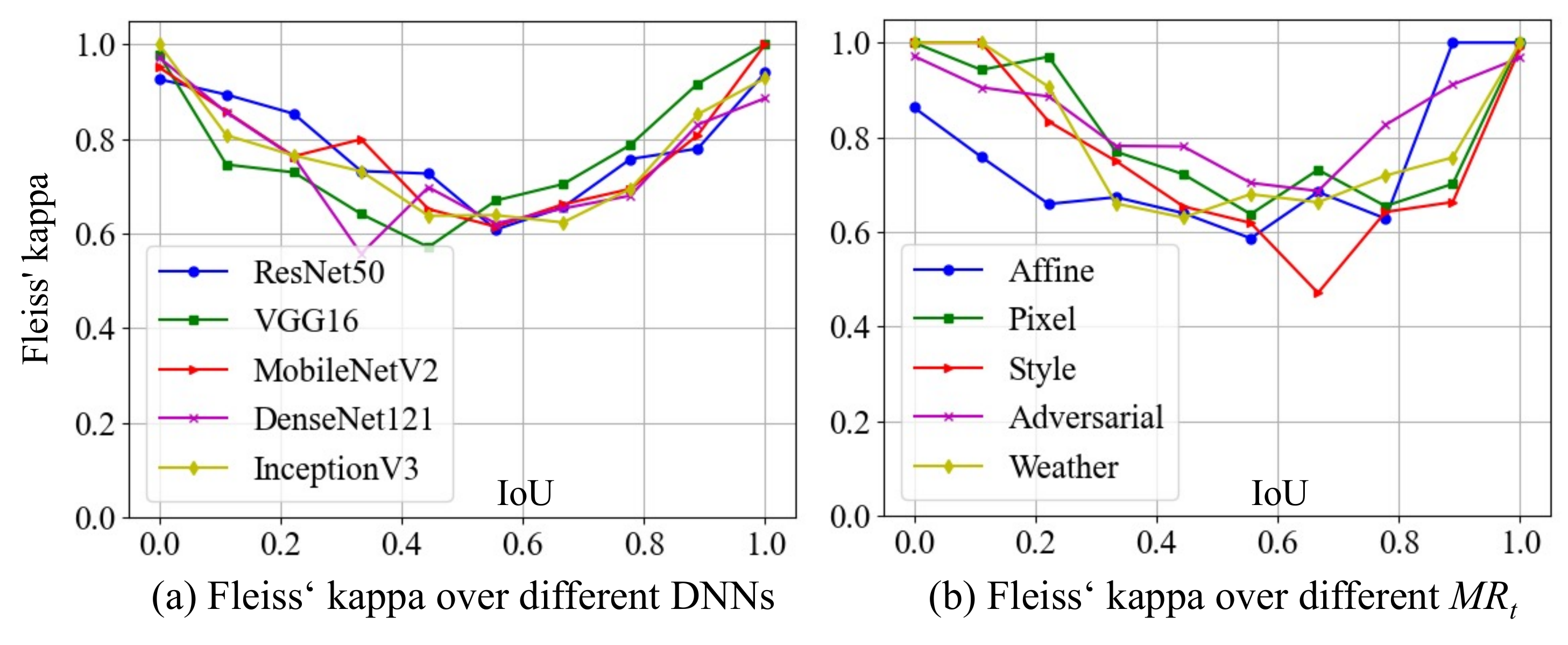}
  \vspace{-20pt}
  \caption{Fleiss' kappa reflects the agreement of participants in
  \S~\ref{subsubsec:pg-student}. We report how the kappa coefficients
  change with the IoU of $\langle i, i' \rangle$ over different DNNs
  and types of \mt. The overall Fleiss' kappa is $0.756$.}
  \label{fig:kappa1}
  \vspace{-5pt}
\end{figure}

\begin{figure}[!ht]
  \centering
  \vspace{-10pt}
  \includegraphics[width=1.01\linewidth]{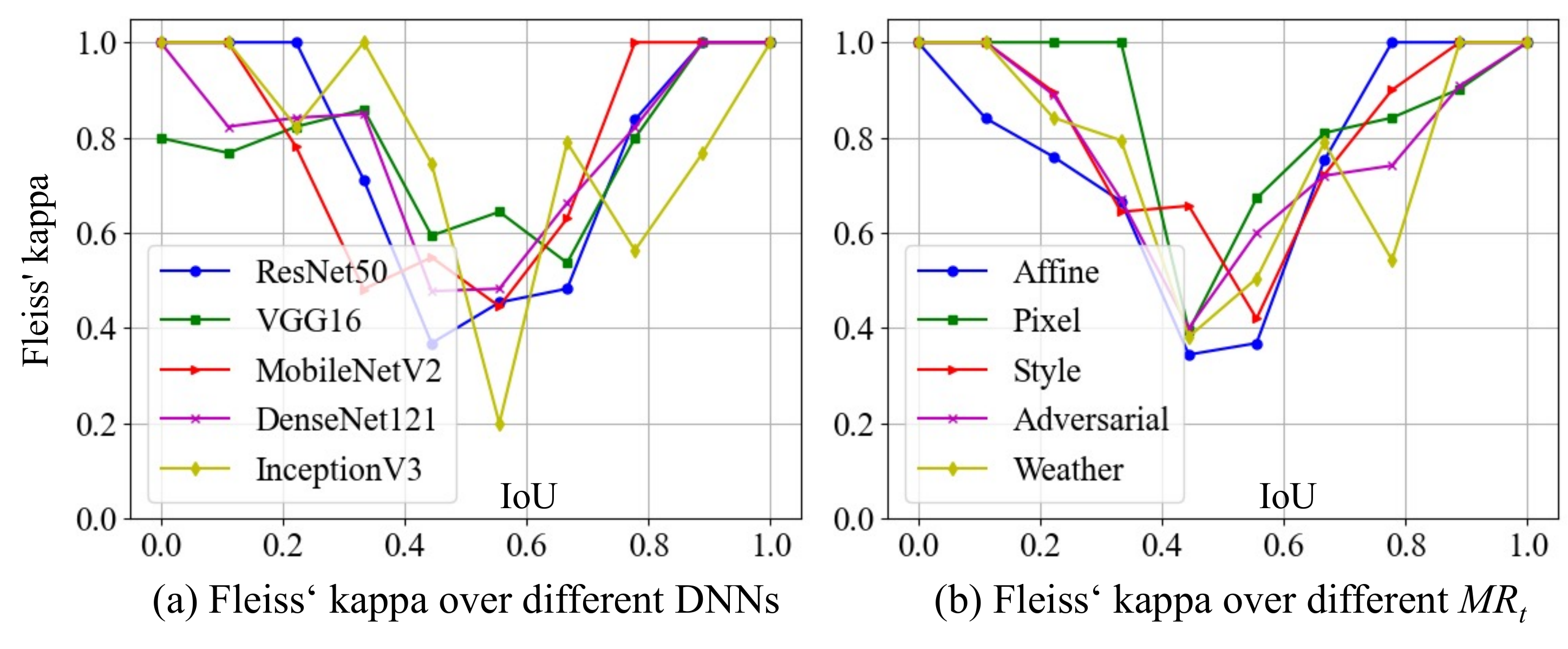}
  \vspace{-20pt}
  \caption{Fleiss' kappa reflects the agreement of participants in
  \S~\ref{subsubsec:non-expert}. We report how the kappa coefficients
  change with the IoU of $\langle i, i' \rangle$ over different DNNs
  and types of \mt. The overall Fleiss' kappa is $0.762$.}
  \label{fig:kappa2}
  \vspace{-5pt}
\end{figure}

We use Fleiss' kappa to evaluate the reliability of our human evaluation
results. The Fleiss' kappa is widely used to assess the agreement among choices
of multiple ($> 2$) participants. A kappa value of $1$ indicates that all
participants totally agree on a choice, whereas kappa values $\leq 0$ imply
participants made completely random choices.

The overall Fleiss' kappa coefficients for the human evaluations in
\S~\ref{subsubsec:pg-student} and \S~\ref{subsubsec:non-expert} are $0.756$ and
$0.762$, indicating a substantial agreement among all participants on all
$\langle i, i' \rangle$ pairs~\cite{gwet2014handbook,sim2005kappa}. We further
present how the kappa coefficients for various DNNs and \mt\ types vary with the
IoU value of decisions $D$ and $D'$ in \F~\ref{fig:kappa1} and
\F~\ref{fig:kappa2}. We note that the results are consistent regardless of the
DNN or \mt: when the IoU values of $\langle i, i' \rangle$'s decisions lie in
$[0, 0.2] \cup [0.8, 1.0]$, the kappa coefficients of all participants are close
to $1$ (and in some cases reach $1$); this indicates a nearly perfect
agreement~\cite{gwet2014handbook}. Overall, all participants concur that two
decisions are different if their corresponding IoU score is less than $0.2$. In sum, we view
the Fleiss' kappa scores are aligned with our observation in
\S~\ref{subsubsec:pg-student}: IoU score threshold 0.2 is generally accurate to
highlight distinct decisions.

\subsubsection{Feedbacks from Participants}
\label{subsubsec:feedback}

We present some representative feedbacks from participants; they are asked to
describe what they primarily rely on to compare decisions. Considering the
following response from a student working on software testing:

\begin{quote}
  ``\textit{I first exclude image pairs that are obviously different, e.g.,
  one color, which should cover a large area, only appear in one of the masked
  image. Then I compare whether the different regions correspond to any
  complete feature, for instance, a whole eye rather than part of
  an eye. I regard two masked images as the same one if their difference
  is only induced by parts of several features.}''
\end{quote}

Notice that he uses a ``process of elimination'' (``\textit{[...] first exclude
[...]}'') to help make comparison. Recall that we give an example of how a DNN
relies on visual concepts to make decisions (``ambulance'') under
\F~\ref{fig:ambulance}. This feedback indicates, to some extent, that the visual
concepts derived by our method are logically reasonable and adequately explain
the DNN decisions. Moreover, considering a response from a student working on
computer vision:

\begin{quote}
  ``\textit{I pay more attention to special textures, for example, 
  marks on plumage of a bird, because I know for most DNNs, they are
  critical for making the prediction.
  }''
\end{quote}

We clarify that these two students' comparison results are mostly
\textit{consistent}. Nevertheless, their strategies and focuses are distinct;
the second participant is attentive to properties like texture. This reflects
the overall difficulty of our research problem, as different experts may base
their comparison on distinct features. It also indicates the high quality of our
extracted decisions, as they are often interpretable by human from multiple
perspectives.

\begin{tcolorbox}[size=small]
  \textbf{Answer to RQ2}: We make several important observation at this step. 1)
  When DNN makes inconsistent predictions over a pair of $\langle i, i'\rangle
  \in \hat{I}_{\neq}$, the decision changes are consistently obvious (maximal
  IoU less than 0.8). 2) When DNN makes consistent predictions over $\langle i,
  i'\rangle \in \hat{I}_{=}$, our human evaluation illustrates that IoU values
  can faithfully reflect the underlying inconsistency of DNN's decisions $D$ and
  $D'$ extracted from $\langle i, i'\rangle \in \hat{I}_{=}$. Empirically, $0.2$
  appears to be a good threshold, such that when the IoU between $D$ and $D'$ is
  below 0.2, $D$ and $D'$ are deemed as different from most human's perspective.
  Participants' agreements are high, though their feedback also indicates the
  difficulty of this research and the high quality of extracted visual concepts.
\end{tcolorbox}

\begin{figure*}
    \captionsetup{skip=5pt}
    \captionsetup[sub]{skip=3pt}
    
    \begin{subfigure}{.19\linewidth}
      \centering
      \resizebox{1.0\linewidth}{!}{

\begin{tikzpicture}
    \begin{axis}[
      ybar,
      bar width=0.2cm,
      grid=major,
      ymin=0,
        yticklabel style={
            /pgf/number format/fixed,
            font=\Large
        },
        xticklabel style={
            font=\Large
        },
      scaled y ticks=false,
      xlabel={\huge IOU},
      ylabel={\huge Hist. value},
      ]
      \addplot + table[x=x, y=y, col sep=&]{fig/tex/txt/resnet50-ImageNet-hist.txt}; \label{hist}
    \end{axis}

    \begin{axis}[
    ylabel near ticks, 
    ylabel style={rotate=180},
    yticklabel pos=right,
    ymin=0,
    axis x line=none,
    ylabel={\huge CDF value},
    legend pos= north west]
    \addlegendimage{ybar interval, ybar interval legend, fill=blue!30, draw=blue}\addlegendentry{\Large Hist.}

    \addplot+[
        red, mark options={scale=0},
        smooth, 
        error bars/.cd, 
            y fixed,
            y dir=both, 
            y explicit
        ] table [x=x, y=y, col sep=&] {fig/tex/txt/resnet50-ImageNet-cdf.txt}; \label{cdf}
        \addlegendimage{/pgfplots/refstyle=size-time}\addlegendentry{\Large CDF}
    \end{axis}

\end{tikzpicture}}
      \caption{\small ResNet \& ImageNet.}
      \label{fig:resnet-imagenet}
    \end{subfigure}
    \begin{subfigure}{.19\linewidth}
      \centering
      \resizebox{1.0\linewidth}{!}{

\begin{tikzpicture}
    \begin{axis}[
      ybar,
      bar width=0.2cm,
      grid=major,
      ymin=0,
        yticklabel style={
            /pgf/number format/fixed,
            font=\Large
        },
        xticklabel style={
            font=\Large
        },
      scaled y ticks=false,
      xlabel={\huge IOU},
      ylabel={\huge Hist. value},
      ]
      \addplot + table[x=x, y=y, col sep=&]{fig/tex/txt/vgg16-bn-ImageNet-hist.txt}; \label{hist}
    \end{axis}

    \begin{axis}[
    ylabel near ticks, 
    ylabel style={rotate=180},
    yticklabel pos=right,
    ymin=0,
    axis x line=none,
    ylabel={\huge CDF value},
    legend pos= north west]
    \addlegendimage{ybar interval, ybar interval legend, fill=blue!30, draw=blue}\addlegendentry{\Large Hist.}

    \addplot+[
        red, mark options={scale=0},
        smooth, 
        error bars/.cd, 
            y fixed,
            y dir=both, 
            y explicit
        ] table [x=x, y=y, col sep=&] {fig/tex/txt/vgg16-bn-ImageNet-cdf.txt}; \label{cdf}
        \addlegendimage{/pgfplots/refstyle=size-time}\addlegendentry{\Large CDF}
    \end{axis}

\end{tikzpicture}}
      \caption{\small VGG \& ImageNet.}
      \label{fig:vgg-imagenet}
    \end{subfigure}
    \begin{subfigure}{.19\linewidth}
      \centering
      \resizebox{1.0\linewidth}{!}{

\begin{tikzpicture}
    \begin{axis}[
      ybar,
      bar width=0.2cm,
      grid=major,
      ymin=0,
        yticklabel style={
            /pgf/number format/fixed,
            font=\Large
        },
        xticklabel style={
            font=\Large
        },
      scaled y ticks=false,
      xlabel={\huge IOU},
      ylabel={\huge Hist. value},
      ]
      \addplot + table[x=x, y=y, col sep=&]{fig/tex/txt/mobilenet-v2-ImageNet-hist.txt}; \label{hist}
    \end{axis}

    \begin{axis}[
    ylabel near ticks, 
    ylabel style={rotate=180},
    yticklabel pos=right,
    ymin=0,
    axis x line=none,
    ylabel={\huge CDF value},
    legend pos= north west]
    \addlegendimage{ybar interval, ybar interval legend, fill=blue!30, draw=blue}\addlegendentry{\Large Hist.}

    \addplot+[
        red, mark options={scale=0},
        smooth, 
        error bars/.cd, 
            y fixed,
            y dir=both, 
            y explicit
        ] table [x=x, y=y, col sep=&] {fig/tex/txt/mobilenet-v2-ImageNet-cdf.txt}; \label{cdf}
        \addlegendimage{/pgfplots/refstyle=size-time}\addlegendentry{\Large CDF}
    \end{axis}

\end{tikzpicture}}
      \caption{\small MobileNet \& ImageNet.}
      \label{fig:mobilenet-imagenet}
    \end{subfigure}
    \begin{subfigure}{.19\linewidth}
      \centering
      \resizebox{1.0\linewidth}{!}{

\begin{tikzpicture}
    \begin{axis}[
      ybar,
      bar width=0.2cm,
      grid=major,
      ymin=0,
        yticklabel style={
            /pgf/number format/fixed,
            font=\Large
        },
        xticklabel style={
            font=\Large
        },
      scaled y ticks=false,
      xlabel={\huge IOU},
      ylabel={\huge Hist. value},
      ]
      \addplot + table[x=x, y=y, col sep=&]{fig/tex/txt/densenet121-ImageNet-hist.txt}; \label{hist}
    \end{axis}

    \begin{axis}[
    ylabel near ticks, 
    ylabel style={rotate=180},
    yticklabel pos=right,
    ymin=0,
    axis x line=none,
    ylabel={\huge CDF value},
    legend pos= north west]
    \addlegendimage{ybar interval, ybar interval legend, fill=blue!30, draw=blue}\addlegendentry{\Large Hist.}

    \addplot+[
        red, mark options={scale=0},
        smooth, 
        error bars/.cd, 
            y fixed,
            y dir=both, 
            y explicit
        ] table [x=x, y=y, col sep=&] {fig/tex/txt/densenet121-ImageNet-cdf.txt}; \label{cdf}
        \addlegendimage{/pgfplots/refstyle=size-time}\addlegendentry{\Large CDF}
    \end{axis}

\end{tikzpicture}}
      \caption{\small DenseNet \& ImageNet.}
      \label{fig:densenet-imagenet}
    \end{subfigure}%
    \begin{subfigure}{.19\linewidth}
      \centering
      \resizebox{1.0\linewidth}{!}{

\begin{tikzpicture}
    \begin{axis}[
      ybar,
      bar width=0.2cm,
      grid=major,
      ymin=0,
        yticklabel style={
            /pgf/number format/fixed,
            font=\Large
        },
        xticklabel style={
            font=\Large
        },
      scaled y ticks=false,
      xlabel={\huge IOU},
      ylabel={\huge Hist. value},
      ]
      \addplot + table[x=x, y=y, col sep=&]{fig/tex/txt/inception-v3-ImageNet-hist.txt}; \label{hist}
    \end{axis}

    \begin{axis}[
    ylabel near ticks, 
    ylabel style={rotate=180},
    yticklabel pos=right,
    ymin=0,
    axis x line=none,
    ylabel={\huge CDF value},
    legend pos= north west]
    \addlegendimage{ybar interval, ybar interval legend, fill=blue!30, draw=blue}\addlegendentry{\Large Hist.}

    \addplot+[
        red, mark options={scale=0},
        smooth, 
        error bars/.cd, 
            y fixed,
            y dir=both, 
            y explicit
        ] table [x=x, y=y, col sep=&] {fig/tex/txt/inception-v3-ImageNet-cdf.txt}; \label{cdf}
        \addlegendimage{/pgfplots/refstyle=size-time}\addlegendentry{\Large CDF}
    \end{axis}

\end{tikzpicture}}
      \caption{\small Inception \& ImageNet.}
      \label{fig:inception-imagenet}
    \end{subfigure}

    \begin{subfigure}{.19\linewidth}
      \centering
      \resizebox{1.0\linewidth}{!}{

\begin{tikzpicture}
    \begin{axis}[
      ybar,
      bar width=0.2cm,
      grid=major,
      ymin=0,
        yticklabel style={
            /pgf/number format/fixed,
            font=\Large
        },
        xticklabel style={
            font=\Large
        },
      scaled y ticks=false,
      xlabel={\huge IOU},
      ylabel={\huge Hist. value},
      ]
      \addplot + table[x=x, y=y, col sep=&]{fig/tex/txt/resnet50-CIFAR10-hist.txt}; \label{hist}
    \end{axis}

    \begin{axis}[
    ylabel near ticks, 
    ylabel style={rotate=180},
    yticklabel pos=right,
    ymin=0,
    axis x line=none,
    ylabel={\huge CDF value},
    legend pos= north west]
    \addlegendimage{ybar interval, ybar interval legend, fill=blue!30, draw=blue}\addlegendentry{\Large Hist.}

    \addplot+[
        red, mark options={scale=0},
        smooth, 
        error bars/.cd, 
            y fixed,
            y dir=both, 
            y explicit
        ] table [x=x, y=y, col sep=&] {fig/tex/txt/resnet50-CIFAR10-cdf.txt}; \label{cdf}
        \addlegendimage{/pgfplots/refstyle=size-time}\addlegendentry{\Large CDF}
    \end{axis}

\end{tikzpicture}}
      \caption{\small ResNet \& CIFAR10.}
      \label{fig:resnet-cifar10}
    \end{subfigure}
    \begin{subfigure}{.19\linewidth}
      \centering
      \resizebox{1.0\linewidth}{!}{

\begin{tikzpicture}
    \begin{axis}[
      ybar,
      bar width=0.2cm,
      grid=major,
      ymin=0,
        yticklabel style={
            /pgf/number format/fixed,
            font=\Large
        },
        xticklabel style={
            font=\Large
        },
      scaled y ticks=false,
      xlabel={\huge IOU},
      ylabel={\huge Hist. value},
      ]
      \addplot + table[x=x, y=y, col sep=&]{fig/tex/txt/vgg16-bn-CIFAR10-hist.txt}; \label{hist}
    \end{axis}

    \begin{axis}[
    ylabel near ticks, 
    ylabel style={rotate=180},
    yticklabel pos=right,
    ymin=0,
    axis x line=none,
    ylabel={\huge CDF value},
    legend pos= north west]
    \addlegendimage{ybar interval, ybar interval legend, fill=blue!30, draw=blue}\addlegendentry{\Large Hist.}

    \addplot+[
        red, mark options={scale=0},
        smooth, 
        error bars/.cd, 
            y fixed,
            y dir=both, 
            y explicit
        ] table [x=x, y=y, col sep=&] {fig/tex/txt/vgg16-bn-CIFAR10-cdf.txt}; \label{cdf}
        \addlegendimage{/pgfplots/refstyle=size-time}\addlegendentry{\Large CDF}
    \end{axis}

\end{tikzpicture}}
      \caption{\small VGG \& CIFAR10.}
      \label{fig:vgg-cifar10}
    \end{subfigure}
    \begin{subfigure}{.19\linewidth}
      \centering
      \resizebox{1.0\linewidth}{!}{

\begin{tikzpicture}
    \begin{axis}[
      ybar,
      bar width=0.2cm,
      grid=major,
      ymin=0,
        yticklabel style={
            /pgf/number format/fixed,
            font=\Large
        },
        xticklabel style={
            font=\Large
        },
      scaled y ticks=false,
      xlabel={\huge IOU},
      ylabel={\huge Hist. value},
      ]
      \addplot + table[x=x, y=y, col sep=&]{fig/tex/txt/mobilenet-v2-CIFAR10-hist.txt}; \label{hist}
    \end{axis}

    \begin{axis}[
    ylabel near ticks, 
    ylabel style={rotate=180},
    yticklabel pos=right,
    ymin=0,
    axis x line=none,
    ylabel={\huge CDF value},
    legend pos= north west]
    \addlegendimage{ybar interval, ybar interval legend, fill=blue!30, draw=blue}\addlegendentry{\Large Hist.}

    \addplot+[
        red, mark options={scale=0},
        smooth, 
        error bars/.cd, 
            y fixed,
            y dir=both, 
            y explicit
        ] table [x=x, y=y, col sep=&] {fig/tex/txt/mobilenet-v2-CIFAR10-cdf.txt}; \label{cdf}
        \addlegendimage{/pgfplots/refstyle=size-time}\addlegendentry{\Large CDF}
    \end{axis}

\end{tikzpicture}}
      \caption{\small MobileNet \& CIFAR10.}
      \label{fig:mobilenet-cifar10}
    \end{subfigure}
    \begin{subfigure}{.19\linewidth}
      \centering
      \resizebox{1.0\linewidth}{!}{

\begin{tikzpicture}
    \begin{axis}[
      ybar,
      bar width=0.2cm,
      grid=major,
      ymin=0,
        yticklabel style={
            /pgf/number format/fixed,
            font=\Large
        },
        xticklabel style={
            font=\Large
        },
      scaled y ticks=false,
      xlabel={\huge IOU},
      ylabel={\huge Hist. value},
      ]
      \addplot + table[x=x, y=y, col sep=&]{fig/tex/txt/LeNet1-MNIST-hist.txt}; \label{hist}
    \end{axis}

    \begin{axis}[
    ylabel near ticks, 
    ylabel style={rotate=180},
    yticklabel pos=right,
    ymin=0,
    axis x line=none,
    ylabel={\huge CDF value},
    legend pos= north west]
    \addlegendimage{ybar interval, ybar interval legend, fill=blue!30, draw=blue}\addlegendentry{\Large Hist.}

    \addplot+[
        red, mark options={scale=0},
        smooth, 
        error bars/.cd, 
            y fixed,
            y dir=both, 
            y explicit
        ] table [x=x, y=y, col sep=&] {fig/tex/txt/LeNet1-MNIST-cdf.txt}; \label{cdf}
        \addlegendimage{/pgfplots/refstyle=size-time}\addlegendentry{\Large CDF}
    \end{axis}

\end{tikzpicture}}
      \caption{\small LeNet1 \& MNIST.}
      \label{fig:lenet1-mnist}
    \end{subfigure}%
    \begin{subfigure}{.19\linewidth}
      \centering
      \resizebox{1.0\linewidth}{!}{

\begin{tikzpicture}
    \begin{axis}[
      ybar,
      bar width=0.2cm,
      grid=major,
      ymin=0,
        yticklabel style={
            /pgf/number format/fixed,
            font=\Large
        },
        xticklabel style={
            font=\Large
        },
      scaled y ticks=false,
      xlabel={\huge IOU},
      ylabel={\huge Hist. value},
      ]
      \addplot + table[x=x, y=y, col sep=&]{fig/tex/txt/LeNet5-MNIST-hist.txt}; \label{hist}
    \end{axis}

    \begin{axis}[
    ylabel near ticks, 
    ylabel style={rotate=180},
    yticklabel pos=right,
    ymin=0,
    axis x line=none,
    ylabel={\huge CDF value},
    legend pos= north west]
    \addlegendimage{ybar interval, ybar interval legend, fill=blue!30, draw=blue}\addlegendentry{\Large Hist.}

    \addplot+[
        red, mark options={scale=0},
        smooth, 
        error bars/.cd, 
            y fixed,
            y dir=both, 
            y explicit
        ] table [x=x, y=y, col sep=&] {fig/tex/txt/LeNet5-MNIST-cdf.txt}; \label{cdf}
        \addlegendimage{/pgfplots/refstyle=size-time}\addlegendentry{\Large CDF}
    \end{axis}

\end{tikzpicture}}
      \caption{\small LeNet5 \& MNIST.}
      \label{fig:lenet5-mnist}
    \end{subfigure}
    
    \vspace{-5pt}
    \caption{Histograms of IoU values (grouped by DNN) over $D$ of $\langle i, i' \rangle \in \hat{I}_{=}$.
    \textcolor{pptred}{Red} line denotes cumulative distribution (CDF).}
    \label{fig:iou-model}
\end{figure*}
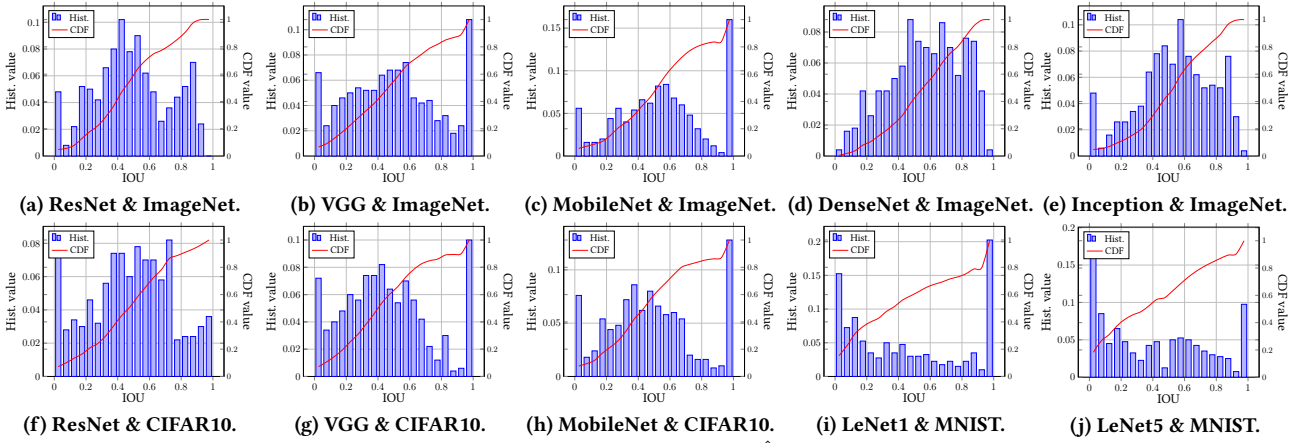

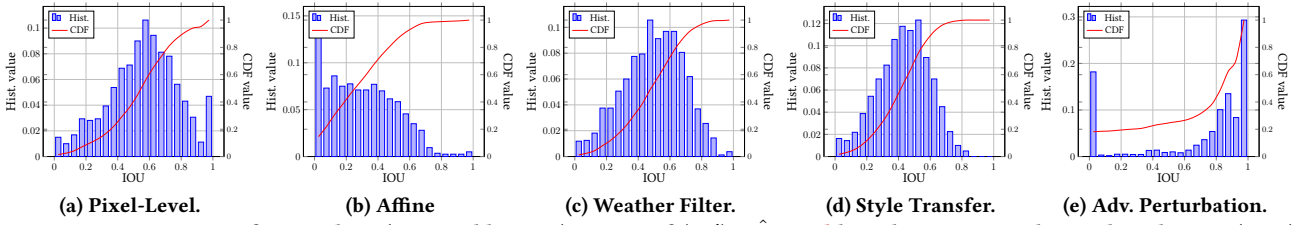
\begin{figure*}
    \captionsetup{skip=5pt}
    \captionsetup[sub]{skip=3pt}
    
    \begin{subfigure}{.19\linewidth}
      \centering
      \resizebox{1.0\linewidth}{!}{

\begin{tikzpicture}
    \begin{axis}[
      ybar,
      bar width=0.2cm,
      grid=major,
      ymin=0,
        yticklabel style={
            /pgf/number format/fixed,
            font=\Large
        },
        xticklabel style={
            font=\Large
        },
      scaled y ticks=false,
      xlabel={\huge IOU},
      ylabel={\huge Hist. value},
      ]
      \addplot + table[x=x, y=y, col sep=&]{fig/tex/txt/P-hist.txt}; \label{hist}
    \end{axis}

    \begin{axis}[
    ylabel near ticks, 
    ylabel style={rotate=180},
    yticklabel pos=right,
    ymin=0,
    axis x line=none,
    ylabel={\huge CDF value},
    legend pos= north west]
    \addlegendimage{ybar interval, ybar interval legend, fill=blue!30, draw=blue}\addlegendentry{\Large Hist.}

    \addplot+[
        red, mark options={scale=0},
        smooth, 
        error bars/.cd, 
            y fixed,
            y dir=both, 
            y explicit
        ] table [x=x, y=y, col sep=&] {fig/tex/txt/P-cdf.txt}; \label{cdf}
        \addlegendimage{/pgfplots/refstyle=size-time}\addlegendentry{\Large CDF}
    \end{axis}

\end{tikzpicture}}
      \caption{\small Pixel-Level.}
      \label{fig:affine}
    \end{subfigure}
    \begin{subfigure}{.19\linewidth}
      \centering
      \resizebox{1.0\linewidth}{!}{

\begin{tikzpicture}
    \begin{axis}[
      ybar,
      bar width=0.2cm,
      grid=major,
      ymin=0,
        yticklabel style={
            /pgf/number format/fixed,
            font=\Large
        },
        xticklabel style={
            font=\Large
        },
      scaled y ticks=false,
      xlabel={\huge IOU},
      ylabel={\huge Hist. value},
      ]
      \addplot + table[x=x, y=y, col sep=&]{fig/tex/txt/G-hist.txt}; \label{hist}
    \end{axis}

    \begin{axis}[
    ylabel near ticks, 
    ylabel style={rotate=180},
    yticklabel pos=right,
    ymin=0,
    axis x line=none,
    ylabel={\huge CDF value},
    legend pos= north west]
    \addlegendimage{ybar interval, ybar interval legend, fill=blue!30, draw=blue}\addlegendentry{\Large Hist.}

    \addplot+[
        red, mark options={scale=0},
        smooth, 
        error bars/.cd, 
            y fixed,
            y dir=both, 
            y explicit
        ] table [x=x, y=y, col sep=&] {fig/tex/txt/G-cdf.txt}; \label{cdf}
        \addlegendimage{/pgfplots/refstyle=size-time}\addlegendentry{\Large CDF}
    \end{axis}

\end{tikzpicture}}
      \caption{\small Affine}
      \label{fig:vgg-imagenet}
    \end{subfigure}
    \begin{subfigure}{.19\linewidth}
      \centering
      \resizebox{1.0\linewidth}{!}{

\begin{tikzpicture}
    \begin{axis}[
      ybar,
      bar width=0.2cm,
      grid=major,
      ymin=0,
        yticklabel style={
            /pgf/number format/fixed,
            font=\Large
        },
        xticklabel style={
            font=\Large
        },
      scaled y ticks=false,
      xlabel={\huge IOU},
      ylabel={\huge Hist. value},
      ]
      \addplot + table[x=x, y=y, col sep=&]{fig/tex/txt/W-hist.txt}; \label{hist}
    \end{axis}

    \begin{axis}[
    ylabel near ticks, 
    ylabel style={rotate=180},
    yticklabel pos=right,
    ymin=0,
    axis x line=none,
    ylabel={\huge CDF value},
    legend pos= north west]
    \addlegendimage{ybar interval, ybar interval legend, fill=blue!30, draw=blue}\addlegendentry{\Large Hist.}

    \addplot+[
        red, mark options={scale=0},
        smooth, 
        error bars/.cd, 
            y fixed,
            y dir=both, 
            y explicit
        ] table [x=x, y=y, col sep=&] {fig/tex/txt/W-cdf.txt}; \label{cdf}
        \addlegendimage{/pgfplots/refstyle=size-time}\addlegendentry{\Large CDF}
    \end{axis}

\end{tikzpicture}}
      \caption{\small Weather Filter.}
      \label{fig:mobilenet-imagenet}
    \end{subfigure}
    \begin{subfigure}{.19\linewidth}
      \centering
      \resizebox{1.0\linewidth}{!}{

\begin{tikzpicture}
    \begin{axis}[
      ybar,
      bar width=0.2cm,
      grid=major,
      ymin=0,
        yticklabel style={
            /pgf/number format/fixed,
            font=\Large
        },
        xticklabel style={
            font=\Large
        },
      scaled y ticks=false,
      xlabel={\huge IOU},
      ylabel={\huge Hist. value},
      ]
      \addplot + table[x=x, y=y, col sep=&]{fig/tex/txt/S-hist.txt}; \label{hist}
    \end{axis}

    \begin{axis}[
    ylabel near ticks, 
    ylabel style={rotate=180},
    yticklabel pos=right,
    ymin=0,
    axis x line=none,
    ylabel={\huge CDF value},
    legend pos= north west]
    \addlegendimage{ybar interval, ybar interval legend, fill=blue!30, draw=blue}\addlegendentry{\Large Hist.}

    \addplot+[
        red, mark options={scale=0},
        smooth, 
        error bars/.cd, 
            y fixed,
            y dir=both, 
            y explicit
        ] table [x=x, y=y, col sep=&] {fig/tex/txt/S-cdf.txt}; \label{cdf}
        \addlegendimage{/pgfplots/refstyle=size-time}\addlegendentry{\Large CDF}
    \end{axis}

\end{tikzpicture}}
      \caption{\small Style Transfer.}
      \label{fig:densenet-imagenet}
    \end{subfigure}%
    \begin{subfigure}{.19\linewidth}
      \centering
      \resizebox{1.0\linewidth}{!}{

\begin{tikzpicture}
    \begin{axis}[
      ybar,
      bar width=0.2cm,
      grid=major,
      ymin=0,
        yticklabel style={
            /pgf/number format/fixed,
            font=\Large
        },
        xticklabel style={
            font=\Large
        },
      scaled y ticks=false,
      xlabel={\huge IOU},
      ylabel={\huge Hist. value},
      ]
      \addplot + table[x=x, y=y, col sep=&]{fig/tex/txt/A-hist.txt}; \label{hist}
    \end{axis}

    \begin{axis}[
    ylabel near ticks, 
    ylabel style={rotate=180},
    yticklabel pos=right,
    ymin=0,
    axis x line=none,
    ylabel={\huge CDF value},
    legend pos= north west]
    \addlegendimage{ybar interval, ybar interval legend, fill=blue!30, draw=blue}\addlegendentry{\Large Hist.}

    \addplot+[
        red, mark options={scale=0},
        smooth, 
        error bars/.cd, 
            y fixed,
            y dir=both, 
            y explicit
        ] table [x=x, y=y, col sep=&] {fig/tex/txt/A-cdf.txt}; \label{cdf}
        \addlegendimage{/pgfplots/refstyle=size-time}\addlegendentry{\Large CDF}
    \end{axis}

\end{tikzpicture}}
      \caption{\small Adv. Perturbation.}
      \label{fig:inception-imagenet}
    \end{subfigure}
    
    \vspace{-5pt}
    \caption{Histograms of IoU values (grouped by \mt) over $D$ of $\langle i, i' \rangle \in \hat{I}_{=}$.
    \textcolor{pptred}{Red} line denotes cumulative distribution (CDF).}
    \label{fig:iou-mt}
\end{figure*}

\subsection{Analysis of Hidden Defects}
\label{subsec:empirical}

This section revisits existing MT-based testing frameworks through the lens of
our decision-based \mr. 
\F~\ref{fig:iou-model} and \F~\ref{fig:iou-mt} show how IoU values, which are
grouped by models or the types of \mt\ over $\langle i, i' \rangle \in
\hat{I}_{=}$, are distributed.

\subsubsection{Results Overview}
\label{subsubsec:quantitative}

We note that for each model, or each type of \mt, almost half of the $\langle i,
i' \rangle$ pairs have IoU values less than $0.5$. Given that DNNs are still
making ``consistent'' predictions over $i$ and $i'$, we thus deem that a large
number of hidden defects are not detected by existing MT frameworks merely
checking DNN prediction consistency. We now analyze hidden defects' characteristics.

\parh{Comparing Different DNNs.}~\F~\ref{fig:iou-model} illustrates that the
distributions of IoU values primarily vary in accordance with the DNN
architectures (e.g., ResNet50 vs. VGG16) rather than with the training datasets
(e.g., CIFAR10 vs. ImageNet). For instance, \F~\ref{fig:iou-model}(a),
\F~\ref{fig:iou-model}(b), and \F~\ref{fig:iou-model}(c) have more distinct
trends, whereas \F~\ref{fig:iou-model}(b) and \F~\ref{fig:iou-model}(g) have
correlated trends.
This is an interesting observation: our evaluated DNNs are all commonly used in
real-world scenarios, and they have different representative structures. Thus,
it is reasonable to assume that, hidden defects due to inconsistent decisions
are mostly induced by different model design. Accordingly, to fix these defects,
developers are expected to focus on DNN architectures rather than simply
enriching the datasets (ImageNet is much more comprehensive than CIFAR10). We
present further discussions on DNN defect repairing in \S~\ref{sec:discussion}.

\parh{Comparing Different \mt.}~\F~\ref{fig:iou-mt} shows that IoU values
corresponds to different \mt\ have distinct distributions.

\parhs{Adversarial Perturbations:}~\F~\ref{fig:iou-mt}(e) shows that the
majority of IoU values are close to two extremes, $0$ and $1$. That is, the
decisions of $\langle i, i' \rangle$, when their corresponding outputs are
identical, are either exactly same or completely different in many cases. It
also has the highest ratio of zero IoU value, indicating its high effectiveness
of flipping DNN predictions. Note that adversarial perturbations are guided by DNN
gradients, which are very informative for describing DNN behaviors. This can
also be reflected from \T~\ref{tab:mt} where key objects in image are
highlighted by adversarial perturbations.
In addition, compared with other \mt\ discussed below, adversarial perturbation
may be more desirable, since it does not introduce much false positive/negative
cases due to challenges in deciding a proper IoU threshold --- as reflected from
our empirical results, users may safely regard two decisions of $\langle i, i'
\rangle$ as inconsistent as long as their IoU value is zero. This saves the
extra effort of human assessments and confirmation.

\parhs{Affine Transformations:}~As in \F~\ref{fig:iou-mt}(b), this \mt\ has the
highest ratio of IoU values less than $0.2$, where the decisions of $\langle i,
i' \rangle$ differ based on our human evaluation. It also has the second-highest
ratio of zero IoU values. Recall we invert an affine transformation on decisions
of $i'$ before computing IoU. Therefore, we may infer that modern DNNs are still
highly sensitive to affine transformations, which can change the position,
viewing angle, size, orientation of objects, thereby introducing more diverse
``patterns'' in images. Compared with other \mt\ (e.g., style transfer and
adversarial perturbation), affine transformations manifest a relative lower cost
and are applicable in most scenarios. It may be reasonable to conclude that
affine transformation is a suitable pre-processing approach for generating
large-scale, diverse images to better stress DNNs. Such a diverse image
collection could be of high benefit in augmenting DNN training data and
repairing DNNs. In fact, this strategy has been adopted by many existing
works~\cite{shorten2019survey}.

\parhs{Style Transfer:}~As in \F~\ref{fig:iou-mt}(d), the highest IoU of this
\mt\ approaches $0.8$. That is, when $i$ is mutated into $i'$ using style
transfer, their corresponding decisions are always distinct (in all our
evaluated cases). Recent works have pointed out that DNNs trained on real
datasets (e.g., ImageNet) have a texture bias: they primarily rely on texture
rather than the shape of objects to make predictions. Accordingly, previous
works proposed training DNNs on style-transferred images to alleviate
texture-bias~\cite{geirhos2018imagenet}. As shown in \T~\ref{tab:mt}, style transfer changes the color
schemes of an image (e.g., from realistic to artistic), which modifies the
texture while retaining the object shape. Our findings illustrate the strength
of style transfer, which, to a great extent, validates the motivation and
argumentation of previous research.
Interestingly, as demonstrated in our evaluation in \S~\ref{subsec:human}, the
highest IoU value calculated over decisions of $\langle i, i' \rangle \in
\hat{I}_{\neq}$ is also close to $0.8$. 
Taking \F~\ref{fig:iou-mt}(d) into account, we hypothesize that an IoU less than
$0.8$ indicates different decisions from the DNN perspective (though are less
human-notable). From our empirical observation, we suggest that safety-critical
DNNs (e.g., autonomous driving) employ an IoU of $0.8$.

\parhs{Pixel-wise Mutation \& Weather Filter:} \F~\ref{fig:iou-mt}(a) and
\F~\ref{fig:iou-mt}(c) have similar IoU value distributions: the majority of IoU
values cluster around $0.5$ and values in $[0, 1]$ are all covered. Both two
\mt\ can be viewed as adding diverse ``noise'' on images. Despite being less
effective than other \mt\ at triggering inconsistent decisions, they still
expose a number of DNN defects. Moreover, it is evident that error-triggering
inputs generated by pixel-wise mutations or weather filters are distinct with
those of style transfer and affine transformations, as they impose different
mutation effects on images. Therefore, all \mt\ proposed by prior
works are demanding to stress DNNs from different angles. Also, since 
pixel-wise mutation and weather filter have lower cost, similar to affine
transformation, it is feasible to generate large-scale, diverse image
collections using these two methods. We thus believe that they are useful for
augmenting DNN training data and repairing DNNs.

\subsubsection{Case Study}
\label{subsubsec:qualitative}
  
As revealed in \F~\ref{fig:iou-model}, taking decision consistency into account
enables detecting a large number of DNN defects from all popular DNNs.
\F~\ref{fig:motivation}, \F~\ref{fig:ambulance}, and \F~\ref{fig:case-same} have
shown several cases, where the decisions in the original images are largely
inconsistent with the mutated images, though DNNs still (incorrectly) yields
consistent labels. Due to limited space, we provide more cases
at~\cite{snapshot}: the inconsistent decisions are mostly induced by both cases
that miss and add visual concepts, as described in \S~\ref{subsec:iou}, and they occur in
images of different classes that are mutated using various methods. Overall, the
diversity indicates that our decision-based oracle can find a broad set of
defects. We will maintain our unveiled cases to benefit follow-up research and
comparison.

\begin{tcolorbox}[size=small]
  \textbf{Answer to RQ3}: We make several key observations. 1)
  Decision-based \mr\ unveils many DNN defects that were overlooked
  by existing MT frameworks. 2) When cross comparing DNNs with different
  architectures or trained using different datasets, we find that decision
  defects primarily root from the DNN architectures. Therefore, repairing the
  exposed defects should focus on tweaking the architectures rather than
  enriching the training data (further discussed in
  \S~\ref{sec:discussion}). 3) When cross comparing different \mt, 
  adversarial perturbations seem more ``reliable'' by omitting fewer
  decision defects. Nevertheless, different \mt\ mutate test inputs from various
  angles. Their exposed defects, when taking decision consistency into account,
  are distinct. This illustrates the necessity of all existing \mt.
\end{tcolorbox}

\begin{table}[t]
  \caption{The number of DNN inferences within one minute on one Nvidia
  GeForce RTX 2080 GPU.}
   \vspace{-10pt}
  \label{tab:cost}
  \centering
\resizebox{0.9\linewidth}{!}{
  \begin{tabular}{c|c|c|c}
    \hline
     & \shortstack{ResNet50\\ImageNet} & \shortstack{ResNet50\\CIFAR10} & \shortstack{LeNet5\\MNIST} \\
    \hline
    Inference w/o decision & $\sim 6,000$/min & $\sim 8,000$/min & $\sim 30,000$/min \\
    \hline
    Inference w/ decision & $\sim 1,600$/min & $\sim 2,000$/min & $\sim 13,000$/min \\
    \hline
  \end{tabular}
  }
  \begin{tablenotes}
    \footnotesize
    \item *ResNet50 is the largest DNN for ImageNet/CIFAR10 in our evaluation.
    \item *LeNet5 is larger than LeNet1.
  \end{tablenotes}
  \vspace{-10pt}
\end{table}

\section{Discussion and Future Work}
\label{sec:discussion}

\noindent \textbf{Cost vs.~Benefit.}~The extra cost of our decision-based oracle
consists of \Circ{a} one backward propagation of \dl, and \Circ{b} the
conversion from prediction-contributing pixels to visual concepts. As clarified
in \S~\ref{subsec:convert}, cost of \Circ{b} is comparable to executing one
extra DNN layer. Given that modern DNNs have tens of layers (e.g., ResNet50 has
50 layers), \Circ{a} introduces the majority of extra cost. We compare DNN
inference speed with and without extracting decisions in \T~\ref{tab:cost}:
extracting decisions slows DNN inference by a factor of four. Since \Circ{b} has
a relatively high tolerance for errors in XAI outputs (recall we abstract XAI
outputs into visual concepts), the overhead can be further reduced by using
various quicker (but possibly less accurate) XAI methods available
off-the-shelf.

From the ``benefit'' perspective, we note that in \S~\ref{sec:evaluation}'s
experiments, only about $1.5\%$ mutated inputs result in incorrect
predictions. $20 \sim 30\%$ of the remaining $98.5\%$ inputs (which induce 
consistent predictions) have IoU values smaller than $0.2$ (i.e., their
decisions are inconsistent), according to the cumulative distribution function
(the red line) in \F~\ref{fig:iou-model} and \F~\ref{fig:iou-mt}. Thus,
taking decision-based oracle into account would reveal many more defects than
before. In sum, given the reasonable cost and the benefit, we believe it is
worthwhile to integrate our new oracle into DNN testing.

\noindent \textbf{Extension to Other Tasks.}~In this research, we primarily
consider images and extract visual concepts from images to form \mr. We clarify
that our work roots the same focus as most works in this field, which test DNNs
for image classification. Nevertheless, from a holistic view, our oracle asserts
if DNN keeps the correct ``focus'' on the seed inputs and the mutated inputs;
our oracle is violated if the focus is changed notably, regardless of the DNN
tasks being evaluated. Thus, our oracle should be applicable to other tasks,
such as activity/emotion recognition where the activity/emotion is recognized
based on visual concepts. 

For common image classification-based tasks, our XAI-based approach can
precisely scope DNN decisions. In complex computer vision scenarios like
auto-driving, classification, localization, and tracking tasks are all involved.
We leave it as one future work to extend our technical pipeline to handle
localization/tracking tasks.
For other tasks where DNNs accept \textit{discrete} inputs (e.g., text, tabular
data, or grid), the minimal semantics-meaningful unit on these inputs is one
discrete unit (similar to an image pixel), such as a word in a sentence, a cell
in a table, or a dot in the grid. Testing such DNNs can be easier, as we may
directly form a decision-based oracle using XAI outputs, including several critical
discrete units.

\noindent \textbf{Threat to Validity.}~This research regards the decisions of a
DNN prediction as a collection of visual concepts, and evaluate the correctness
of decisions by removing/keeping all visual concepts together. One threat
is that some identified visual concepts are incorrect. In practice, it is
challenging to assess the correctness of each visual concept individually,
because removing/keeping part of the identified visual concepts may break the
integrity of the decisions and lead to a different decision process.

However, we clarify that a few incorrect visual concepts, if they exist at all,
should not affect our decision-based oracle for two reasons. First, \dl\ is a
well-established XAI tool whose errors are pixel-wise, typically in the form of
several inaccurately estimated contribution scores. However, instead of using
the exact scores, we simply recognize pixels with positive scores. Moreover, the
procedure of abstracting pixel-level contributions to visual concepts can also
reduce inaccuracy. Second, we do not regard two decisions as inconsistent based
solely on deviations in visual concepts. Instead, we deem decisions are
``inconsistent'' when their overlapping is below a small threshold. Overall, the
threat of potentially incorrect XAI outputs is mostly eliminated in our
implementation.

\section{Conclusion}
\label{sec:conclusion}

To reveal hidden DNN defects due to ill-decisions, this paper proposes to
extend MT-based DNN testing by checking DNN decision consistency. 
Our evaluation shows that decision-based MT exhibits promising
detectability for DNN defects. Our findings can provide
insights for researchers that aim to launch MT toward DNNs.

\begin{acks}                            
   We thank the anonymous ASE reviewers and our shepherd for their valuable feedback
   and constructive comments.
   We also thank all participants of our human evaluation.
\end{acks}

\bibliographystyle{ACM-Reference-Format}
\bibliography{bib/main}


\begin{thebibliography}{69}


\ifx \showCODEN    \undefined \def \showCODEN     #1{\unskip}     \fi
\ifx \showDOI      \undefined \def \showDOI       #1{#1}\fi
\ifx \showISBNx    \undefined \def \showISBNx     #1{\unskip}     \fi
\ifx \showISBNxiii \undefined \def \showISBNxiii  #1{\unskip}     \fi
\ifx \showISSN     \undefined \def \showISSN      #1{\unskip}     \fi
\ifx \showLCCN     \undefined \def \showLCCN      #1{\unskip}     \fi
\ifx \shownote     \undefined \def \shownote      #1{#1}          \fi
\ifx \showarticletitle \undefined \def \showarticletitle #1{#1}   \fi
\ifx \showURL      \undefined \def \showURL       {\relax}        \fi
\providecommand\bibfield[2]{#2}
\providecommand\bibinfo[2]{#2}
\providecommand\natexlab[1]{#1}
\providecommand\showeprint[2][]{arXiv:#2}

\bibitem[\protect\citeauthoryear{??}{sna}{[n.d.]}]%
        {snapshot}
 \bibinfo{year}{[n.d.]}\natexlab{}.
\newblock \bibinfo{title}{Research Artifact}.
\newblock
  \bibinfo{howpublished}{\url{https://github.com/Yuanyuan-Yuan/Decision-Oracle}}.
\newblock


\bibitem[\protect\citeauthoryear{Ancona, Oztireli, and Gross}{Ancona
  et~al\mbox{.}}{2019}]%
        {ancona2019explaining}
\bibfield{author}{\bibinfo{person}{Marco Ancona}, \bibinfo{person}{Cengiz
  Oztireli}, {and} \bibinfo{person}{Markus Gross}.}
  \bibinfo{year}{2019}\natexlab{}.
\newblock \showarticletitle{Explaining deep neural networks with a polynomial
  time for shapley value approximation} \emph{(\bibinfo{series}{PMLR})}.
\newblock


\bibitem[\protect\citeauthoryear{Bach, Binder, Montavon, Klauschen, M{\"u}ller,
  and Samek}{Bach et~al\mbox{.}}{2015}]%
        {bach2015pixel}
\bibfield{author}{\bibinfo{person}{Sebastian Bach}, \bibinfo{person}{Alexander
  Binder}, \bibinfo{person}{Gr{\'e}goire Montavon}, \bibinfo{person}{Frederick
  Klauschen}, \bibinfo{person}{Klaus-Robert M{\"u}ller}, {and}
  \bibinfo{person}{Wojciech Samek}.} \bibinfo{year}{2015}\natexlab{}.
\newblock \showarticletitle{On pixel-wise explanations for non-linear
  classifier decisions by layer-wise relevance propagation}.
\newblock \bibinfo{journal}{\emph{PloS one}} \bibinfo{volume}{10},
  \bibinfo{number}{7} (\bibinfo{year}{2015}), \bibinfo{pages}{e0130140}.
\newblock


\bibitem[\protect\citeauthoryear{Carlini and Wagner}{Carlini and
  Wagner}{2017}]%
        {carlini2017towards}
\bibfield{author}{\bibinfo{person}{Nicholas Carlini} {and}
  \bibinfo{person}{David Wagner}.} \bibinfo{year}{2017}\natexlab{}.
\newblock \showarticletitle{Towards evaluating the robustness of neural
  networks} \emph{(\bibinfo{series}{IEEE SP})}.
\newblock


\bibitem[\protect\citeauthoryear{Carter, Armstrong, Schubert, Johnson, and
  Olah}{Carter et~al\mbox{.}}{2019}]%
        {carter2019exploring}
\bibfield{author}{\bibinfo{person}{Shan Carter}, \bibinfo{person}{Zan
  Armstrong}, \bibinfo{person}{Ludwig Schubert}, \bibinfo{person}{Ian Johnson},
  {and} \bibinfo{person}{Chris Olah}.} \bibinfo{year}{2019}\natexlab{}.
\newblock \showarticletitle{Exploring neural networks with activation atlases}.
\newblock \bibinfo{journal}{\emph{Distill}}  \bibinfo{volume}{1}
  (\bibinfo{year}{2019}), \bibinfo{pages}{2}.
\newblock


\bibitem[\protect\citeauthoryear{Chattopadhay, Sarkar, Howlader, and
  Balasubramanian}{Chattopadhay et~al\mbox{.}}{2018}]%
        {chattopadhay2018grad}
\bibfield{author}{\bibinfo{person}{Aditya Chattopadhay},
  \bibinfo{person}{Anirban Sarkar}, \bibinfo{person}{Prantik Howlader}, {and}
  \bibinfo{person}{Vineeth~N Balasubramanian}.}
  \bibinfo{year}{2018}\natexlab{}.
\newblock \showarticletitle{Grad-cam++: Generalized gradient-based visual
  explanations for deep convolutional networks}
  \emph{(\bibinfo{series}{WACV})}.
\newblock


\bibitem[\protect\citeauthoryear{Chen, Cheung, and Yiu}{Chen
  et~al\mbox{.}}{1998}]%
        {chen1998metamorphic}
\bibfield{author}{\bibinfo{person}{Tsong~Y Chen}, \bibinfo{person}{Shing~C
  Cheung}, {and} \bibinfo{person}{Shiu~Ming Yiu}.}
  \bibinfo{year}{1998}\natexlab{}.
\newblock \bibinfo{booktitle}{\emph{Metamorphic testing: a new approach for
  generating next test cases}}.
\newblock \bibinfo{type}{{T}echnical {R}eport}. \bibinfo{institution}{Technical
  Report HKUST-CS98-01, Department of Computer Science, Hong Kong~…}.
\newblock


\bibitem[\protect\citeauthoryear{Covert and Lee}{Covert and Lee}{2021}]%
        {covert2021improving}
\bibfield{author}{\bibinfo{person}{Ian Covert} {and} \bibinfo{person}{Su-In
  Lee}.} \bibinfo{year}{2021}\natexlab{}.
\newblock \showarticletitle{Improving KernelSHAP: Practical Shapley value
  estimation using linear regression} \emph{(\bibinfo{series}{ICAIS})}.
\newblock


\bibitem[\protect\citeauthoryear{Datta, Sen, and Zick}{Datta
  et~al\mbox{.}}{2016}]%
        {datta2016algorithmic}
\bibfield{author}{\bibinfo{person}{Anupam Datta}, \bibinfo{person}{Shayak Sen},
  {and} \bibinfo{person}{Yair Zick}.} \bibinfo{year}{2016}\natexlab{}.
\newblock \showarticletitle{Algorithmic transparency via quantitative input
  influence: Theory and experiments with learning systems}. In
  \bibinfo{booktitle}{\emph{IEEE SP}}.
\newblock


\bibitem[\protect\citeauthoryear{Demir, Eniser, and Sen}{Demir
  et~al\mbox{.}}{2019}]%
        {demir2019deepsmartfuzzer}
\bibfield{author}{\bibinfo{person}{Samet Demir}, \bibinfo{person}{Hasan~Ferit
  Eniser}, {and} \bibinfo{person}{Alper Sen}.} \bibinfo{year}{2019}\natexlab{}.
\newblock \showarticletitle{{DeepSmartFuzzer}: Reward Guided Test Generation
  For Deep Learning}.
\newblock \bibinfo{journal}{\emph{arXiv preprint arXiv:1911.10621}}
  (\bibinfo{year}{2019}).
\newblock


\bibitem[\protect\citeauthoryear{Deng, Dong, Socher, Li, Li, and Fei-Fei}{Deng
  et~al\mbox{.}}{2009}]%
        {imagenet}
\bibfield{author}{\bibinfo{person}{J. Deng}, \bibinfo{person}{W. Dong},
  \bibinfo{person}{R. Socher}, \bibinfo{person}{L.-J. Li}, \bibinfo{person}{K.
  Li}, {and} \bibinfo{person}{L. Fei-Fei}.} \bibinfo{year}{2009}\natexlab{}.
\newblock \showarticletitle{{ImageNet: A Large-Scale Hierarchical Image
  Database}}. In \bibinfo{booktitle}{\emph{CVPR09}}.
\newblock


\bibitem[\protect\citeauthoryear{Deng}{Deng}{2012}]%
        {deng2012mnist}
\bibfield{author}{\bibinfo{person}{Li Deng}.} \bibinfo{year}{2012}\natexlab{}.
\newblock \showarticletitle{The mnist database of handwritten digit images for
  machine learning research [best of the web]}.
\newblock \bibinfo{journal}{\emph{IEEE Signal Processing Magazine}}
  (\bibinfo{year}{2012}).
\newblock


\bibitem[\protect\citeauthoryear{Dwarakanath, Ahuja, Podder, Vinu, Naskar, and
  Koushik}{Dwarakanath et~al\mbox{.}}{2019}]%
        {dwarakanath2019metamorphic}
\bibfield{author}{\bibinfo{person}{Anurag Dwarakanath}, \bibinfo{person}{Manish
  Ahuja}, \bibinfo{person}{Sanjay Podder}, \bibinfo{person}{Silja Vinu},
  \bibinfo{person}{Arijit Naskar}, {and} \bibinfo{person}{MV Koushik}.}
  \bibinfo{year}{2019}\natexlab{}.
\newblock \showarticletitle{Metamorphic testing of a deep learning based
  forecaster}. In \bibinfo{booktitle}{\emph{MET}}.
\newblock


\bibitem[\protect\citeauthoryear{Dwarakanath, Ahuja, Sikand, Rao, Bose, Dubash,
  and Podder}{Dwarakanath et~al\mbox{.}}{2018}]%
        {dwarakanath2018identifying}
\bibfield{author}{\bibinfo{person}{Anurag Dwarakanath}, \bibinfo{person}{Manish
  Ahuja}, \bibinfo{person}{Samarth Sikand}, \bibinfo{person}{Raghotham~M. Rao},
  \bibinfo{person}{R.~P. Jagadeesh~Chandra Bose}, \bibinfo{person}{Neville
  Dubash}, {and} \bibinfo{person}{Sanjay Podder}.}
  \bibinfo{year}{2018}\natexlab{}.
\newblock \showarticletitle{Identifying Implementation Bugs in Machine Learning
  Based Image Classifiers Using Metamorphic Testing}. In
  \bibinfo{booktitle}{\emph{ISSTA}}.
\newblock


\bibitem[\protect\citeauthoryear{Geirhos, Rubisch, Michaelis, Bethge, Wichmann,
  and Brendel}{Geirhos et~al\mbox{.}}{2018}]%
        {geirhos2018imagenet}
\bibfield{author}{\bibinfo{person}{Robert Geirhos}, \bibinfo{person}{Patricia
  Rubisch}, \bibinfo{person}{Claudio Michaelis}, \bibinfo{person}{Matthias
  Bethge}, \bibinfo{person}{Felix~A Wichmann}, {and} \bibinfo{person}{Wieland
  Brendel}.} \bibinfo{year}{2018}\natexlab{}.
\newblock \showarticletitle{ImageNet-trained CNNs are biased towards texture;
  increasing shape bias improves accuracy and robustness}.
\newblock \bibinfo{journal}{\emph{arXiv preprint arXiv:1811.12231}}
  (\bibinfo{year}{2018}).
\newblock


\bibitem[\protect\citeauthoryear{Girshick}{Girshick}{2015}]%
        {girshick2015fast}
\bibfield{author}{\bibinfo{person}{Ross Girshick}.}
  \bibinfo{year}{2015}\natexlab{}.
\newblock \showarticletitle{Fast r-cnn}. In
  \bibinfo{booktitle}{\emph{Proceedings of the IEEE international conference on
  computer vision}}. \bibinfo{pages}{1440--1448}.
\newblock


\bibitem[\protect\citeauthoryear{Goodfellow, Shlens, and Szegedy}{Goodfellow
  et~al\mbox{.}}{2015}]%
        {goodfellow2014explaining}
\bibfield{author}{\bibinfo{person}{Ian~J Goodfellow}, \bibinfo{person}{Jonathon
  Shlens}, {and} \bibinfo{person}{Christian Szegedy}.}
  \bibinfo{year}{2015}\natexlab{}.
\newblock \showarticletitle{Explaining and harnessing adversarial examples}. In
  \bibinfo{booktitle}{\emph{ICLR}}.
\newblock


\bibitem[\protect\citeauthoryear{Gunning, Stefik, Choi, Miller, Stumpf, and
  Yang}{Gunning et~al\mbox{.}}{2019}]%
        {gunning2019xai}
\bibfield{author}{\bibinfo{person}{David Gunning}, \bibinfo{person}{Mark
  Stefik}, \bibinfo{person}{Jaesik Choi}, \bibinfo{person}{Timothy Miller},
  \bibinfo{person}{Simone Stumpf}, {and} \bibinfo{person}{Guang-Zhong Yang}.}
  \bibinfo{year}{2019}\natexlab{}.
\newblock \showarticletitle{XAI—Explainable artificial intelligence}.
\newblock \bibinfo{journal}{\emph{Science Robotics}} \bibinfo{volume}{4},
  \bibinfo{number}{37} (\bibinfo{year}{2019}), \bibinfo{pages}{eaay7120}.
\newblock


\bibitem[\protect\citeauthoryear{Gwet}{Gwet}{2014}]%
        {gwet2014handbook}
\bibfield{author}{\bibinfo{person}{Kilem~L Gwet}.}
  \bibinfo{year}{2014}\natexlab{}.
\newblock \bibinfo{booktitle}{\emph{Handbook of inter-rater reliability: The
  definitive guide to measuring the extent of agreement among raters}}.
\newblock \bibinfo{publisher}{Advanced Analytics, LLC}.
\newblock


\bibitem[\protect\citeauthoryear{He, Gkioxari, Doll{\'a}r, and Girshick}{He
  et~al\mbox{.}}{2017}]%
        {he2017mask}
\bibfield{author}{\bibinfo{person}{Kaiming He}, \bibinfo{person}{Georgia
  Gkioxari}, \bibinfo{person}{Piotr Doll{\'a}r}, {and} \bibinfo{person}{Ross
  Girshick}.} \bibinfo{year}{2017}\natexlab{}.
\newblock \showarticletitle{{Mask R-CNN}}. In \bibinfo{booktitle}{\emph{CVPR}}.
  \bibinfo{pages}{2961--2969}.
\newblock


\bibitem[\protect\citeauthoryear{He, Zhang, Ren, and Sun}{He
  et~al\mbox{.}}{2016}]%
        {he2016resnet}
\bibfield{author}{\bibinfo{person}{Kaiming He}, \bibinfo{person}{Xiangyu
  Zhang}, \bibinfo{person}{Shaoqing Ren}, {and} \bibinfo{person}{Jian Sun}.}
  \bibinfo{year}{2016}\natexlab{}.
\newblock \showarticletitle{Deep residual learning for image recognition}. In
  \bibinfo{booktitle}{\emph{CVPR}}. \bibinfo{pages}{770--778}.
\newblock


\bibitem[\protect\citeauthoryear{Hein, Andriushchenko, and Bitterwolf}{Hein
  et~al\mbox{.}}{2019}]%
        {hein2019relu}
\bibfield{author}{\bibinfo{person}{Matthias Hein}, \bibinfo{person}{Maksym
  Andriushchenko}, {and} \bibinfo{person}{Julian Bitterwolf}.}
  \bibinfo{year}{2019}\natexlab{}.
\newblock \showarticletitle{Why relu networks yield high-confidence predictions
  far away from the training data and how to mitigate the problem}. In
  \bibinfo{booktitle}{\emph{CVPR}}.
\newblock


\bibitem[\protect\citeauthoryear{Howard, Zhu, Chen, Kalenichenko, Wang, Weyand,
  Andreetto, and Adam}{Howard et~al\mbox{.}}{2017}]%
        {howard2017mobilenets}
\bibfield{author}{\bibinfo{person}{Andrew~G Howard}, \bibinfo{person}{Menglong
  Zhu}, \bibinfo{person}{Bo Chen}, \bibinfo{person}{Dmitry Kalenichenko},
  \bibinfo{person}{Weijun Wang}, \bibinfo{person}{Tobias Weyand},
  \bibinfo{person}{Marco Andreetto}, {and} \bibinfo{person}{Hartwig Adam}.}
  \bibinfo{year}{2017}\natexlab{}.
\newblock \showarticletitle{Mobilenets: Efficient convolutional neural networks
  for mobile vision applications}.
\newblock \bibinfo{journal}{\emph{arXiv preprint arXiv:1704.04861}}
  (\bibinfo{year}{2017}).
\newblock


\bibitem[\protect\citeauthoryear{Hu, Marsso, Czarnecki, Salay, Shen, and
  Chechik}{Hu et~al\mbox{.}}{2022}]%
        {hu2022if}
\bibfield{author}{\bibinfo{person}{Boyue~Caroline Hu}, \bibinfo{person}{Lina
  Marsso}, \bibinfo{person}{Krzysztof Czarnecki}, \bibinfo{person}{Rick Salay},
  \bibinfo{person}{Huakun Shen}, {and} \bibinfo{person}{Marsha Chechik}.}
  \bibinfo{year}{2022}\natexlab{}.
\newblock \showarticletitle{If a Human Can See It, So Should Your System:
  Reliability Requirements for Machine Vision Components}.
\newblock \bibinfo{journal}{\emph{arXiv preprint arXiv:2202.03930}}
  (\bibinfo{year}{2022}).
\newblock


\bibitem[\protect\citeauthoryear{Huang, Liu, Van Der~Maaten, and
  Weinberger}{Huang et~al\mbox{.}}{2017}]%
        {huang2017densely}
\bibfield{author}{\bibinfo{person}{Gao Huang}, \bibinfo{person}{Zhuang Liu},
  \bibinfo{person}{Laurens Van Der~Maaten}, {and} \bibinfo{person}{Kilian~Q
  Weinberger}.} \bibinfo{year}{2017}\natexlab{}.
\newblock \showarticletitle{Densely connected convolutional networks}. In
  \bibinfo{booktitle}{\emph{CVPR}}.
\newblock


\bibitem[\protect\citeauthoryear{Jung, Wada, Crall, Tanaka, Graving, Reinders,
  Yadav, Banerjee, Vecsei, Kraft, Rui, Borovec, Vallentin, Zhydenko, Pfeiffer,
  Cook, Fernández, De~Rainville, Weng, Ayala-Acevedo, Meudec, Laporte,
  et~al\mbox{.}}{Jung et~al\mbox{.}}{2020}]%
        {imgaug}
\bibfield{author}{\bibinfo{person}{Alexander~B. Jung}, \bibinfo{person}{Kentaro
  Wada}, \bibinfo{person}{Jon Crall}, \bibinfo{person}{Satoshi Tanaka},
  \bibinfo{person}{Jake Graving}, \bibinfo{person}{Christoph Reinders},
  \bibinfo{person}{Sarthak Yadav}, \bibinfo{person}{Joy Banerjee},
  \bibinfo{person}{Gábor Vecsei}, \bibinfo{person}{Adam Kraft},
  \bibinfo{person}{Zheng Rui}, \bibinfo{person}{Jirka Borovec},
  \bibinfo{person}{Christian Vallentin}, \bibinfo{person}{Semen Zhydenko},
  \bibinfo{person}{Kilian Pfeiffer}, \bibinfo{person}{Ben Cook},
  \bibinfo{person}{Ismael Fernández}, \bibinfo{person}{François-Michel
  De~Rainville}, \bibinfo{person}{Chi-Hung Weng}, \bibinfo{person}{Abner
  Ayala-Acevedo}, \bibinfo{person}{Raphael Meudec}, \bibinfo{person}{Matias
  Laporte}, {et~al\mbox{.}}} \bibinfo{year}{2020}\natexlab{}.
\newblock \bibinfo{title}{{imgaug}}.
\newblock \bibinfo{howpublished}{\url{https://github.com/aleju/imgaug}}.
\newblock
\newblock
\shownote{Online; accessed 01-Feb-2020.}


\bibitem[\protect\citeauthoryear{Krizhevsky, Hinton, et~al\mbox{.}}{Krizhevsky
  et~al\mbox{.}}{2009}]%
        {krizhevsky2009cifar}
\bibfield{author}{\bibinfo{person}{Alex Krizhevsky}, \bibinfo{person}{Geoffrey
  Hinton}, {et~al\mbox{.}}} \bibinfo{year}{2009}\natexlab{}.
\newblock \showarticletitle{Learning multiple layers of features from tiny
  images}.
\newblock  (\bibinfo{year}{2009}).
\newblock


\bibitem[\protect\citeauthoryear{Kundel and Nodine}{Kundel and Nodine}{1983}]%
        {kundel1983visual}
\bibfield{author}{\bibinfo{person}{HL Kundel} {and} \bibinfo{person}{CF
  Nodine}.} \bibinfo{year}{1983}\natexlab{}.
\newblock \showarticletitle{A visual concept shapes image perception.}
\newblock \bibinfo{journal}{\emph{Radiology}} \bibinfo{volume}{146},
  \bibinfo{number}{2} (\bibinfo{year}{1983}), \bibinfo{pages}{363--368}.
\newblock


\bibitem[\protect\citeauthoryear{Kurakin, Goodfellow, and Bengio}{Kurakin
  et~al\mbox{.}}{2016}]%
        {kurakin2016adversarial}
\bibfield{author}{\bibinfo{person}{Alexey Kurakin}, \bibinfo{person}{Ian
  Goodfellow}, {and} \bibinfo{person}{Samy Bengio}.}
  \bibinfo{year}{2016}\natexlab{}.
\newblock \showarticletitle{Adversarial examples in the physical world}.
\newblock \bibinfo{journal}{\emph{arXiv preprint arXiv:1607.02533}}
  (\bibinfo{year}{2016}).
\newblock


\bibitem[\protect\citeauthoryear{LeCun, Boser, Denker, Henderson, Howard,
  Hubbard, and Jackel}{LeCun et~al\mbox{.}}{1989a}]%
        {lecun1989handwritten}
\bibfield{author}{\bibinfo{person}{Yann LeCun}, \bibinfo{person}{Bernhard
  Boser}, \bibinfo{person}{John Denker}, \bibinfo{person}{Donnie Henderson},
  \bibinfo{person}{Richard Howard}, \bibinfo{person}{Wayne Hubbard}, {and}
  \bibinfo{person}{Lawrence Jackel}.} \bibinfo{year}{1989}\natexlab{a}.
\newblock \showarticletitle{Handwritten digit recognition with a
  back-propagation network}.
\newblock \bibinfo{journal}{\emph{NIPS}} (\bibinfo{year}{1989}).
\newblock


\bibitem[\protect\citeauthoryear{LeCun, Boser, Denker, Henderson, Howard,
  Hubbard, and Jackel}{LeCun et~al\mbox{.}}{1989b}]%
        {lecun1989backpropagation}
\bibfield{author}{\bibinfo{person}{Yann LeCun}, \bibinfo{person}{Bernhard
  Boser}, \bibinfo{person}{John~S Denker}, \bibinfo{person}{Donnie Henderson},
  \bibinfo{person}{Richard~E Howard}, \bibinfo{person}{Wayne Hubbard}, {and}
  \bibinfo{person}{Lawrence~D Jackel}.} \bibinfo{year}{1989}\natexlab{b}.
\newblock \showarticletitle{Backpropagation applied to handwritten zip code
  recognition}.
\newblock \bibinfo{journal}{\emph{Neural computation}} \bibinfo{volume}{1},
  \bibinfo{number}{4} (\bibinfo{year}{1989}), \bibinfo{pages}{541--551}.
\newblock


\bibitem[\protect\citeauthoryear{LeCun, Bottou, Bengio, and Haffner}{LeCun
  et~al\mbox{.}}{1998}]%
        {lecun1998gradient}
\bibfield{author}{\bibinfo{person}{Yann LeCun}, \bibinfo{person}{L{\'e}on
  Bottou}, \bibinfo{person}{Yoshua Bengio}, {and} \bibinfo{person}{Patrick
  Haffner}.} \bibinfo{year}{1998}\natexlab{}.
\newblock \showarticletitle{Gradient-based learning applied to document
  recognition}.
\newblock \bibinfo{journal}{\emph{Proc. IEEE}} \bibinfo{volume}{86},
  \bibinfo{number}{11} (\bibinfo{year}{1998}), \bibinfo{pages}{2278--2324}.
\newblock


\bibitem[\protect\citeauthoryear{Li, Wang, Liu, Wang, Wang, and Gao}{Li
  et~al\mbox{.}}{2022}]%
        {li2022cctest}
\bibfield{author}{\bibinfo{person}{Zongjie Li}, \bibinfo{person}{Chaozheng
  Wang}, \bibinfo{person}{Zhibo Liu}, \bibinfo{person}{Haoxuan Wang},
  \bibinfo{person}{Shuai Wang}, {and} \bibinfo{person}{Cuiyun Gao}.}
  \bibinfo{year}{2022}\natexlab{}.
\newblock \showarticletitle{CCTEST: Testing and Repairing Code Completion
  Systems}.
\newblock \bibinfo{journal}{\emph{arXiv preprint arXiv:2208.08289}}
  (\bibinfo{year}{2022}).
\newblock


\bibitem[\protect\citeauthoryear{Linardatos, Papastefanopoulos, and
  Kotsiantis}{Linardatos et~al\mbox{.}}{2020}]%
        {linardatos2020explainable}
\bibfield{author}{\bibinfo{person}{Pantelis Linardatos},
  \bibinfo{person}{Vasilis Papastefanopoulos}, {and} \bibinfo{person}{Sotiris
  Kotsiantis}.} \bibinfo{year}{2020}\natexlab{}.
\newblock \showarticletitle{Explainable ai: A review of machine learning
  interpretability methods}.
\newblock \bibinfo{journal}{\emph{Entropy}} \bibinfo{volume}{23},
  \bibinfo{number}{1} (\bibinfo{year}{2020}), \bibinfo{pages}{18}.
\newblock


\bibitem[\protect\citeauthoryear{Lundberg and Lee}{Lundberg and Lee}{2017}]%
        {lundberg2017unified}
\bibfield{author}{\bibinfo{person}{Scott~M Lundberg} {and}
  \bibinfo{person}{Su-In Lee}.} \bibinfo{year}{2017}\natexlab{}.
\newblock \showarticletitle{A unified approach to interpreting model
  predictions}.
\newblock \bibinfo{journal}{\emph{Advances in neural information processing
  systems}}  \bibinfo{volume}{30} (\bibinfo{year}{2017}).
\newblock


\bibitem[\protect\citeauthoryear{Ma, Zhang, Sun, Xue, Li, Juefei-Xu, Xie, Li,
  Liu, Zhao, et~al\mbox{.}}{Ma et~al\mbox{.}}{2018}]%
        {ma2018deepmutation}
\bibfield{author}{\bibinfo{person}{Lei Ma}, \bibinfo{person}{Fuyuan Zhang},
  \bibinfo{person}{Jiyuan Sun}, \bibinfo{person}{Minhui Xue},
  \bibinfo{person}{Bo Li}, \bibinfo{person}{Felix Juefei-Xu},
  \bibinfo{person}{Chao Xie}, \bibinfo{person}{Li Li}, \bibinfo{person}{Yang
  Liu}, \bibinfo{person}{Jianjun Zhao}, {et~al\mbox{.}}}
  \bibinfo{year}{2018}\natexlab{}.
\newblock \showarticletitle{Deepmutation: Mutation testing of deep learning
  systems}. In \bibinfo{booktitle}{\emph{ISSRE}}.
\newblock


\bibitem[\protect\citeauthoryear{Ma and Wang}{Ma and Wang}{2021}]%
        {ma2021mt}
\bibfield{author}{\bibinfo{person}{Pingchuan Ma} {and} \bibinfo{person}{Shuai
  Wang}.} \bibinfo{year}{2021}\natexlab{}.
\newblock \showarticletitle{MT-teql: evaluating and augmenting neural NLIDB on
  real-world linguistic and schema variations}.
\newblock  (\bibinfo{year}{2021}).
\newblock


\bibitem[\protect\citeauthoryear{Ma, Wang, and Liu}{Ma et~al\mbox{.}}{2020}]%
        {ma2020metamorphic}
\bibfield{author}{\bibinfo{person}{Pingchuan Ma}, \bibinfo{person}{Shuai Wang},
  {and} \bibinfo{person}{Jin Liu}.} \bibinfo{year}{2020}\natexlab{}.
\newblock \showarticletitle{Metamorphic Testing and Certified Mitigation of
  Fairness Violations in NLP Models}. In \bibinfo{booktitle}{\emph{IJCAI}}.
  \bibinfo{pages}{458--465}.
\newblock


\bibitem[\protect\citeauthoryear{Madry, Makelov, Schmidt, Tsipras, and
  Vladu}{Madry et~al\mbox{.}}{2017}]%
        {madry2017towards}
\bibfield{author}{\bibinfo{person}{Aleksander Madry},
  \bibinfo{person}{Aleksandar Makelov}, \bibinfo{person}{Ludwig Schmidt},
  \bibinfo{person}{Dimitris Tsipras}, {and} \bibinfo{person}{Adrian Vladu}.}
  \bibinfo{year}{2017}\natexlab{}.
\newblock \showarticletitle{Towards deep learning models resistant to
  adversarial attacks}.
\newblock \bibinfo{journal}{\emph{arXiv preprint arXiv:1706.06083}}
  (\bibinfo{year}{2017}).
\newblock


\bibitem[\protect\citeauthoryear{Mao, Wei, Yang, Wang, Huang, and Yuille}{Mao
  et~al\mbox{.}}{2015}]%
        {mao2015learning}
\bibfield{author}{\bibinfo{person}{Junhua Mao}, \bibinfo{person}{Xu Wei},
  \bibinfo{person}{Yi Yang}, \bibinfo{person}{Jiang Wang},
  \bibinfo{person}{Zhiheng Huang}, {and} \bibinfo{person}{Alan~L Yuille}.}
  \bibinfo{year}{2015}\natexlab{}.
\newblock \showarticletitle{Learning like a child: Fast novel visual concept
  learning from sentence descriptions of images}. In
  \bibinfo{booktitle}{\emph{ICCV}}.
\newblock


\bibitem[\protect\citeauthoryear{Montavon, Binder, Lapuschkin, Samek, and
  M{\"u}ller}{Montavon et~al\mbox{.}}{2019}]%
        {montavon2019layer}
\bibfield{author}{\bibinfo{person}{Gr{\'e}goire Montavon},
  \bibinfo{person}{Alexander Binder}, \bibinfo{person}{Sebastian Lapuschkin},
  \bibinfo{person}{Wojciech Samek}, {and} \bibinfo{person}{Klaus-Robert
  M{\"u}ller}.} \bibinfo{year}{2019}\natexlab{}.
\newblock \showarticletitle{Layer-wise relevance propagation: an overview}.
\newblock \bibinfo{journal}{\emph{Explainable AI: interpreting, explaining and
  visualizing deep learning}} (\bibinfo{year}{2019}).
\newblock


\bibitem[\protect\citeauthoryear{Nakajima and Chen}{Nakajima and Chen}{2019}]%
        {nakajima2019generating}
\bibfield{author}{\bibinfo{person}{Shin Nakajima} {and}
  \bibinfo{person}{Tsong~Yueh Chen}.} \bibinfo{year}{2019}\natexlab{}.
\newblock \showarticletitle{Generating biased dataset for metamorphic testing
  of machine learning programs}. In \bibinfo{booktitle}{\emph{IFIP-ICTSS}}.
\newblock


\bibitem[\protect\citeauthoryear{Nguyen, Yosinski, and Clune}{Nguyen
  et~al\mbox{.}}{2015}]%
        {nguyen2015deep}
\bibfield{author}{\bibinfo{person}{Anh Nguyen}, \bibinfo{person}{Jason
  Yosinski}, {and} \bibinfo{person}{Jeff Clune}.}
  \bibinfo{year}{2015}\natexlab{}.
\newblock \showarticletitle{Deep neural networks are easily fooled: High
  confidence predictions for unrecognizable images}. In
  \bibinfo{booktitle}{\emph{Proceedings of the IEEE conference on computer
  vision and pattern recognition}}. \bibinfo{pages}{427--436}.
\newblock


\bibitem[\protect\citeauthoryear{Odena and Goodfellow}{Odena and
  Goodfellow}{2018}]%
        {odena2018tensorfuzz}
\bibfield{author}{\bibinfo{person}{Augustus Odena} {and} \bibinfo{person}{Ian
  Goodfellow}.} \bibinfo{year}{2018}\natexlab{}.
\newblock \showarticletitle{{Tensorfuzz}: Debugging neural networks with
  coverage-guided fuzzing}.
\newblock \bibinfo{journal}{\emph{arXiv preprint arXiv:1807.10875}}
  (\bibinfo{year}{2018}).
\newblock


\bibitem[\protect\citeauthoryear{Otsu}{Otsu}{1979}]%
        {otsu1979threshold}
\bibfield{author}{\bibinfo{person}{Nobuyuki Otsu}.}
  \bibinfo{year}{1979}\natexlab{}.
\newblock \showarticletitle{A threshold selection method from gray-level
  histograms}.
\newblock \bibinfo{journal}{\emph{IEEE transactions on systems, man, and
  cybernetics}} \bibinfo{volume}{9}, \bibinfo{number}{1}
  (\bibinfo{year}{1979}), \bibinfo{pages}{62--66}.
\newblock


\bibitem[\protect\citeauthoryear{Pei, Cao, Yang, and Jana}{Pei
  et~al\mbox{.}}{2017}]%
        {pei2017deepxplore}
\bibfield{author}{\bibinfo{person}{Kexin Pei}, \bibinfo{person}{Yinzhi Cao},
  \bibinfo{person}{Junfeng Yang}, {and} \bibinfo{person}{Suman Jana}.}
  \bibinfo{year}{2017}\natexlab{}.
\newblock \showarticletitle{{DeepXplore}: Automated Whitebox Testing of Deep
  Learning Systems} \emph{(\bibinfo{series}{SOSP '17})}.
\newblock


\bibitem[\protect\citeauthoryear{Ren, He, Girshick, and Sun}{Ren
  et~al\mbox{.}}{2015}]%
        {ren2015faster}
\bibfield{author}{\bibinfo{person}{Shaoqing Ren}, \bibinfo{person}{Kaiming He},
  \bibinfo{person}{Ross Girshick}, {and} \bibinfo{person}{Jian Sun}.}
  \bibinfo{year}{2015}\natexlab{}.
\newblock \showarticletitle{Faster r-cnn: Towards real-time object detection
  with region proposal networks}. In \bibinfo{booktitle}{\emph{Advances in
  neural information processing systems}}. \bibinfo{pages}{91--99}.
\newblock


\bibitem[\protect\citeauthoryear{Ribeiro, Singh, and Guestrin}{Ribeiro
  et~al\mbox{.}}{2016}]%
        {ribeiro2016should}
\bibfield{author}{\bibinfo{person}{Marco~Tulio Ribeiro},
  \bibinfo{person}{Sameer Singh}, {and} \bibinfo{person}{Carlos Guestrin}.}
  \bibinfo{year}{2016}\natexlab{}.
\newblock \showarticletitle{" Why should i trust you?" Explaining the
  predictions of any classifier}. In \bibinfo{booktitle}{\emph{KDD}}.
\newblock


\bibitem[\protect\citeauthoryear{Segura, Fraser, Sanchez, and
  Ruiz-Cort{\'e}s}{Segura et~al\mbox{.}}{2016}]%
        {segura2016survey}
\bibfield{author}{\bibinfo{person}{Sergio Segura}, \bibinfo{person}{Gordon
  Fraser}, \bibinfo{person}{Ana~B Sanchez}, {and} \bibinfo{person}{Antonio
  Ruiz-Cort{\'e}s}.} \bibinfo{year}{2016}\natexlab{}.
\newblock \showarticletitle{A survey on metamorphic testing}.
\newblock \bibinfo{journal}{\emph{IEEE TSE}} (\bibinfo{year}{2016}).
\newblock


\bibitem[\protect\citeauthoryear{Selvaraju, Cogswell, Das, Vedantam, Parikh,
  and Batra}{Selvaraju et~al\mbox{.}}{2017}]%
        {selvaraju2017grad}
\bibfield{author}{\bibinfo{person}{Ramprasaath~R Selvaraju},
  \bibinfo{person}{Michael Cogswell}, \bibinfo{person}{Abhishek Das},
  \bibinfo{person}{Ramakrishna Vedantam}, \bibinfo{person}{Devi Parikh}, {and}
  \bibinfo{person}{Dhruv Batra}.} \bibinfo{year}{2017}\natexlab{}.
\newblock \showarticletitle{Grad-cam: Visual explanations from deep networks
  via gradient-based localization}. In \bibinfo{booktitle}{\emph{ICCV}}.
\newblock


\bibitem[\protect\citeauthoryear{Shapley}{Shapley}{2016}]%
        {shapley201617}
\bibfield{author}{\bibinfo{person}{Lloyd~S Shapley}.}
  \bibinfo{year}{2016}\natexlab{}.
\newblock \bibinfo{booktitle}{\emph{A value for n-person games}}.
\newblock \bibinfo{publisher}{Princeton University Press}.
\newblock


\bibitem[\protect\citeauthoryear{Shorten and Khoshgoftaar}{Shorten and
  Khoshgoftaar}{2019}]%
        {shorten2019survey}
\bibfield{author}{\bibinfo{person}{Connor Shorten} {and}
  \bibinfo{person}{Taghi~M Khoshgoftaar}.} \bibinfo{year}{2019}\natexlab{}.
\newblock \showarticletitle{A survey on image data augmentation for deep
  learning}.
\newblock \bibinfo{journal}{\emph{Journal of big data}} \bibinfo{volume}{6},
  \bibinfo{number}{1} (\bibinfo{year}{2019}), \bibinfo{pages}{1--48}.
\newblock


\bibitem[\protect\citeauthoryear{Shrikumar, Greenside, and Kundaje}{Shrikumar
  et~al\mbox{.}}{2017}]%
        {shrikumar2017learning}
\bibfield{author}{\bibinfo{person}{Avanti Shrikumar}, \bibinfo{person}{Peyton
  Greenside}, {and} \bibinfo{person}{Anshul Kundaje}.}
  \bibinfo{year}{2017}\natexlab{}.
\newblock \showarticletitle{Learning important features through propagating
  activation differences}. In \bibinfo{booktitle}{\emph{ICML}}.
\newblock


\bibitem[\protect\citeauthoryear{Sim and Wright}{Sim and Wright}{2005}]%
        {sim2005kappa}
\bibfield{author}{\bibinfo{person}{Julius Sim} {and} \bibinfo{person}{Chris~C
  Wright}.} \bibinfo{year}{2005}\natexlab{}.
\newblock \showarticletitle{The kappa statistic in reliability studies: use,
  interpretation, and sample size requirements}.
\newblock \bibinfo{journal}{\emph{Physical therapy}} (\bibinfo{year}{2005}).
\newblock


\bibitem[\protect\citeauthoryear{Simonyan and Zisserman}{Simonyan and
  Zisserman}{2014}]%
        {simonyan2014very}
\bibfield{author}{\bibinfo{person}{Karen Simonyan} {and}
  \bibinfo{person}{Andrew Zisserman}.} \bibinfo{year}{2014}\natexlab{}.
\newblock \showarticletitle{Very deep convolutional networks for large-scale
  image recognition}.
\newblock \bibinfo{journal}{\emph{arXiv preprint arXiv:1409.1556}}
  (\bibinfo{year}{2014}).
\newblock


\bibitem[\protect\citeauthoryear{Sundararajan, Taly, and Yan}{Sundararajan
  et~al\mbox{.}}{2017}]%
        {sundararajan2017axiomatic}
\bibfield{author}{\bibinfo{person}{Mukund Sundararajan}, \bibinfo{person}{Ankur
  Taly}, {and} \bibinfo{person}{Qiqi Yan}.} \bibinfo{year}{2017}\natexlab{}.
\newblock \showarticletitle{Axiomatic attribution for deep networks}. In
  \bibinfo{booktitle}{\emph{ICML}}.
\newblock


\bibitem[\protect\citeauthoryear{Szegedy, Vanhoucke, Ioffe, Shlens, and
  Wojna}{Szegedy et~al\mbox{.}}{2016}]%
        {szegedy2016rethinking}
\bibfield{author}{\bibinfo{person}{Christian Szegedy}, \bibinfo{person}{Vincent
  Vanhoucke}, \bibinfo{person}{Sergey Ioffe}, \bibinfo{person}{Jon Shlens},
  {and} \bibinfo{person}{Zbigniew Wojna}.} \bibinfo{year}{2016}\natexlab{}.
\newblock \showarticletitle{Rethinking the inception architecture for computer
  vision}. In \bibinfo{booktitle}{\emph{CVPR}}.
\newblock


\bibitem[\protect\citeauthoryear{Tian, Ma, Wen, Liu, Cheung, and Zhang}{Tian
  et~al\mbox{.}}{2021}]%
        {tian2021extent}
\bibfield{author}{\bibinfo{person}{Yongqiang Tian}, \bibinfo{person}{Shiqing
  Ma}, \bibinfo{person}{Ming Wen}, \bibinfo{person}{Yepang Liu},
  \bibinfo{person}{Shing-Chi Cheung}, {and} \bibinfo{person}{Xiangyu Zhang}.}
  \bibinfo{year}{2021}\natexlab{}.
\newblock \showarticletitle{To what extent do DNN-based image classification
  models make unreliable inferences?}
\newblock \bibinfo{journal}{\emph{Empirical Software Engineering}}
  \bibinfo{volume}{26}, \bibinfo{number}{5} (\bibinfo{year}{2021}),
  \bibinfo{pages}{1--40}.
\newblock


\bibitem[\protect\citeauthoryear{Tian, Pei, Jana, and Ray}{Tian
  et~al\mbox{.}}{2018}]%
        {tian2018deeptest}
\bibfield{author}{\bibinfo{person}{Yuchi Tian}, \bibinfo{person}{Kexin Pei},
  \bibinfo{person}{Suman Jana}, {and} \bibinfo{person}{Baishakhi Ray}.}
  \bibinfo{year}{2018}\natexlab{}.
\newblock \showarticletitle{{DeepTest}: Automated Testing of
  Deep-neural-network-driven Autonomous Cars} \emph{(\bibinfo{series}{ICSE
  '18})}.
\newblock


\bibitem[\protect\citeauthoryear{Van~de Sande, Gevers, and Snoek}{Van~de Sande
  et~al\mbox{.}}{2008}]%
        {van2008comparison}
\bibfield{author}{\bibinfo{person}{Koen~EA Van~de Sande}, \bibinfo{person}{Theo
  Gevers}, {and} \bibinfo{person}{Cees~GM Snoek}.}
  \bibinfo{year}{2008}\natexlab{}.
\newblock \showarticletitle{A comparison of color features for visual concept
  classification}. In \bibinfo{booktitle}{\emph{Proceedings of the 2008
  international conference on Content-based image and video retrieval}}.
  \bibinfo{pages}{141--150}.
\newblock


\bibitem[\protect\citeauthoryear{Wang, Xu, Xu, Ma, and Lu}{Wang
  et~al\mbox{.}}{2020}]%
        {wang2020dissector}
\bibfield{author}{\bibinfo{person}{Huiyan Wang}, \bibinfo{person}{Jingwei Xu},
  \bibinfo{person}{Chang Xu}, \bibinfo{person}{Xiaoxing Ma}, {and}
  \bibinfo{person}{Jian Lu}.} \bibinfo{year}{2020}\natexlab{}.
\newblock \showarticletitle{Dissector: Input validation for deep learning
  applications by layer dissection} \emph{(\bibinfo{series}{ICSE})}.
\newblock


\bibitem[\protect\citeauthoryear{Wang and Su}{Wang and Su}{2020}]%
        {wang2020metamorphic}
\bibfield{author}{\bibinfo{person}{Shuai Wang} {and} \bibinfo{person}{Zhendong
  Su}.} \bibinfo{year}{2020}\natexlab{}.
\newblock \showarticletitle{Metamorphic Object Insertion for Testing Object
  Detection Systems}. In \bibinfo{booktitle}{\emph{ASE}}.
\newblock


\bibitem[\protect\citeauthoryear{Xiao, Liu, Yuan, Pang, and Wang}{Xiao
  et~al\mbox{.}}{2022}]%
        {xiao2022metamorphic}
\bibfield{author}{\bibinfo{person}{Dongwei Xiao}, \bibinfo{person}{Zhibo Liu},
  \bibinfo{person}{Yuanyuan Yuan}, \bibinfo{person}{Qi Pang}, {and}
  \bibinfo{person}{Shuai Wang}.} \bibinfo{year}{2022}\natexlab{}.
\newblock \showarticletitle{Metamorphic Testing of Deep Learning Compilers}.
\newblock  (\bibinfo{year}{2022}).
\newblock


\bibitem[\protect\citeauthoryear{Xie, Ma, Juefei-Xu, Chen, Xue, Li, Liu, Zhao,
  Yin, and See}{Xie et~al\mbox{.}}{2018}]%
        {xie2018coverage}
\bibfield{author}{\bibinfo{person}{Xiaofei Xie}, \bibinfo{person}{Lei Ma},
  \bibinfo{person}{Felix Juefei-Xu}, \bibinfo{person}{Hongxu Chen},
  \bibinfo{person}{Minhui Xue}, \bibinfo{person}{Bo Li}, \bibinfo{person}{Yang
  Liu}, \bibinfo{person}{Jianjun Zhao}, \bibinfo{person}{Jianxiong Yin}, {and}
  \bibinfo{person}{Simon See}.} \bibinfo{year}{2018}\natexlab{}.
\newblock \showarticletitle{Coverage-guided fuzzing for deep neural networks}.
\newblock \bibinfo{journal}{\emph{arXiv preprint arXiv:1809.01266}}
  (\bibinfo{year}{2018}).
\newblock


\bibitem[\protect\citeauthoryear{Yuan, Pang, and Wang}{Yuan
  et~al\mbox{.}}{2021a}]%
        {yuan2021enhancing}
\bibfield{author}{\bibinfo{person}{Yuanyuan Yuan}, \bibinfo{person}{Qi Pang},
  {and} \bibinfo{person}{Shuai Wang}.} \bibinfo{year}{2021}\natexlab{a}.
\newblock \showarticletitle{Enhancing Deep Neural Networks Testing by
  Traversing Data Manifold}.
\newblock \bibinfo{journal}{\emph{arXiv preprint arXiv:2112.01956}}
  (\bibinfo{year}{2021}).
\newblock


\bibitem[\protect\citeauthoryear{Yuan, Pang, and Wang}{Yuan
  et~al\mbox{.}}{2022}]%
        {yuan2022unveiling}
\bibfield{author}{\bibinfo{person}{Yuanyuan Yuan}, \bibinfo{person}{Qi Pang},
  {and} \bibinfo{person}{Shuai Wang}.} \bibinfo{year}{2022}\natexlab{}.
\newblock \showarticletitle{Unveiling Hidden DNN Defects with Decision-Based
  Metamorphic Testing} \emph{(\bibinfo{series}{ASE})}.
\newblock


\bibitem[\protect\citeauthoryear{Yuan, Wang, Jiang, and Chen}{Yuan
  et~al\mbox{.}}{2021b}]%
        {yuan2021perception}
\bibfield{author}{\bibinfo{person}{Yuanyuan Yuan}, \bibinfo{person}{Shuai
  Wang}, \bibinfo{person}{Mingyue Jiang}, {and} \bibinfo{person}{Tsong~Yueh
  Chen}.} \bibinfo{year}{2021}\natexlab{b}.
\newblock \showarticletitle{Perception Matters: Detecting Perception Failures
  of VQA Models Using Metamorphic Testing}. In
  \bibinfo{booktitle}{\emph{CVPR}}.
\newblock


\bibitem[\protect\citeauthoryear{Zhang, Harman, Ma, and Liu}{Zhang
  et~al\mbox{.}}{2020}]%
        {zhang2020machine}
\bibfield{author}{\bibinfo{person}{Jie~M Zhang}, \bibinfo{person}{Mark Harman},
  \bibinfo{person}{Lei Ma}, {and} \bibinfo{person}{Yang Liu}.}
  \bibinfo{year}{2020}\natexlab{}.
\newblock \showarticletitle{Machine learning testing: Survey, landscapes and
  horizons}.
\newblock \bibinfo{journal}{\emph{TSE}} (\bibinfo{year}{2020}).
\newblock


\bibitem[\protect\citeauthoryear{Zhang, Zhang, Zhang, Liu, and Khurshid}{Zhang
  et~al\mbox{.}}{2018}]%
        {zhang2018deeproad}
\bibfield{author}{\bibinfo{person}{Mengshi Zhang}, \bibinfo{person}{Yuqun
  Zhang}, \bibinfo{person}{Lingming Zhang}, \bibinfo{person}{Cong Liu}, {and}
  \bibinfo{person}{Sarfraz Khurshid}.} \bibinfo{year}{2018}\natexlab{}.
\newblock \showarticletitle{{DeepRoad: GAN-based Metamorphic Testing and Input
  Validation Framework for Autonomous Driving Systems}}. In
  \bibinfo{booktitle}{\emph{ASE}}.
\newblock


\end{thebibliography}

\end{document}